\documentclass[11pt]{article}

\usepackage{mathpazo, mathtools,amssymb}
\usepackage{fullpage, stmaryrd, bm}
\usepackage[dvipsnames]{xcolor}
\usepackage[T1]{fontenc}
\usepackage{setspace, todonotes}

\usepackage{amsthm,amsmath}
\usepackage{thmtools,thm-restate}

\numberwithin{equation}{section}
\declaretheoremstyle[bodyfont=\it,qed=\qedsymbol]{noproofstyle}

\declaretheorem[numberlike=equation]{observation}

\declaretheorem[name=Observation,numbered=no]{observation*}

\declaretheorem[numberlike=equation]{fact}

\declaretheorem[numberlike=equation]{theorem}

\declaretheorem[name=Theorem,numbered=no]{theorem*}

\declaretheorem[numberlike=equation]{lemma}
\declaretheorem[name=Lemma,numbered=no]{lemma*}

\declaretheorem[name=Corollary,numbered=no]{corollary*}

\declaretheorem[numberlike=equation]{proposition}
\declaretheorem[name=Proposition,numbered=no]{proposition*}

\declaretheorem[name=Claim,numbered=no]{claim*}

\declaretheorem[name=Conjecture,numbered=no]{conjecture*}

\declaretheorem[name=Question,numbered=no]{question*}

\declaretheoremstyle[bodyfont=\it,qed=$\lozenge$]{defstyle} 

\declaretheorem[numberlike=equation,style=defstyle]{definition}
\declaretheorem[unnumbered,name=Definition,style=defstyle]{definition*}

\declaretheorem[unnumbered,name=Example,style=defstyle]{example*}

\declaretheorem[numberlike=equation,style=defstyle]{notation}
\declaretheorem[unnumbered,name=Notation=defstyle]{notation*}

\declaretheorem[unnumbered,name=Construction,style=defstyle]{construction*}

\declaretheorem[numberlike=equation,style=defstyle]{remark}
\declaretheorem[unnumbered,name=Remark,style=defstyle]{remark*}

\renewcommand{\phi}{\varphi}
\renewcommand{\epsilon}{\varepsilon}

\newcommand{\ckt}{\textbf{$\mathcal{C}$}}

\newcommand{\abp}{\textbf{$\mathcal{A}$}}
\newcommand{\roabp}{\mathsf{ROABP}}

\newcommand{\coeff}{\operatorname{\mathsf{coeff}}}
\newcommand{\supp}{\operatorname{\mathsf{supp}}}
\newcommand{\mult}{\operatorname{\mathsf{mult}}}
\def\rank{\operatorname{rank}}
\newcommand{\Char}{\mbox{\rm Char}}
\newcommand{\ips}{\mathsf{IPS}}
\newcommand{\Lin}{\mathsf{Lin}}
\newcommand{\multips}{\mathsf{mult\text{-}IPS}}

\newcommand{\algrank}{\mathsf{alg\text{-}rank}}
\def\poly{\operatorname{poly}}

\renewcommand{\mod}[1]{\ (\mathrm{mod}\ #1)}

\newcommand{\LM}{\mbox{\small\rm LM}} 
\newcommand{\TL}{\mbox{\small  \rm TM}}
\newcommand{\eqdef}{:=}

\newcommand{\eval}{\mbox{\small\rm Eval-dim}}
\newcommand{\mon}{\mbox{\small\rm mon}}

\DeclareMathOperator{\var}{\mbox{\small\rm var}}

\newcommand{\esym}[2]{\mathrm{ESYM}_{#1,#2}}

\def\veca{\bar{a}}
\def\vecb{\bar{b}}
\def\vect{\bar{t}}

\def\vecx{\bar{x}}
\def\vecy{\bar{y}}
\def\vecz{\bar{z}}
\def\set#1{\left\{ #1 \right\}}
\def\inparen#1{\left( #1 \right)}
\def\inbrace#1{\left\{ #1 \right\}}
\def\setdef#1#2{\inbrace{ #1 \ : \ #2}}   
\def\abs#1{\left| #1 \right|}
\def\insquare#1{\left[ #1 \right]}

\def\VBP{\mathsf{VBP}}

\def\VNP{\mathsf{VNP}}

\usepackage{nth}
\usepackage{intcalc}
\usepackage{etoolbox}
\usepackage{xstring}

\usepackage{ifpdf}
\ifpdf
\else
\usepackage[quadpoints=false]{hypdvips}
\fi

\newcommand{\ECCCurl}[2]{http://eccc.hpi-web.de/report/\ifnumcomp{#1}{>}{93}{19}{20}#1/#2/}

\newcommand{\shortECCC}[2]{\texttt{\href{http://eccc.hpi-web.de/report/\ifnumcomp{#1}{>}{93}{19}{20}#1/#2/}{eccc:TR#1-#2}}}

\newcommand{\parseECCC}[1]{
\StrSubstitute{#1}{TR}{}[\tmpstring]%
\IfSubStr{\tmpstring}{/}{ 
\StrBefore{\tmpstring}{/}[\ecccyear]%
\StrBehind{\tmpstring}{/}[\ecccreport]%
}{
\StrBefore{\tmpstring}{-}[\ecccyear]%
\StrBehind{\tmpstring}{-}[\ecccreport]%
}%
\shortECCC{\ecccyear}{\ecccreport}}

\RequirePackage[colorlinks=true]{hyperref}
\hypersetup{
  linkcolor=[rgb]{0.1,0.5,0.5},
  citecolor=[rgb]{0.5, 0.1, 0.5},
  urlcolor=[rgb]{0.5, 0.1, 0.5}
}

\newif\ifcomment

\newcommand{\prerona}[1]{\ifcomment{\color{blue} [{\bf Prerona:} #1]}\fi}

\commenttrue 

\title{$\ips$ Lower Bounds for Formulas and Sum of ROABPs}
\author{Prerona Chatterjee\thanks{School of Computer Sciences, NISER, Bhubaneswar, India. Email: \texttt{prerona.ch@gmail.com}} \quad Utsab Ghosal\thanks{Chennai Mathematical Institute, Chennai, India. Partially supported by Infosys grant. Email : \texttt{ghosal@cmi.ac.in}.} \quad Partha Mukhopadhyay\thanks{Chennai Mathematical Institute, Chennai, India. Partially supported by Infosys grant. Email : \texttt{partham@cmi.ac.in}.} \quad Amit Sinhababu\thanks{Chennai Mathematical Institute, Chennai, India. Partially supported by Infosys grant. Email : \texttt{amitks@cmi.ac.in}.}}
\date{\today}

\begin{document}

\maketitle

\begin{abstract}
We give new lower bounds for the fragments of the Ideal Proof System ($\ips$) introduced by Grochow and Pitassi \cite{GP18}. The Ideal Proof System is a central topic in algebraic proof complexity developed in the context of Nullstellensatz refutation \cite{BIKPP94} and simulates Extended Frege efficiently. Our main results are as follows.

\begin{itemize}
\item $\multips_{\Lin'}$: We prove nearly quadratic-size formula lower bound for multilinear refutation (over the Boolean hypercube) of a variant of the subset-sum axiom polynomial. Extending this, we obtain a nearly matching qualitative statement for a constant degree target polynomial. 
\item $\ips_{\Lin'}:$ Over the fields of characteristic zero, we prove exponential-size sum-of-$\roabp$s lower bound for the refutation of a variant of the subset-sum axiom polynomial. The result also extends over the fields of positive characteristics when the target polynomial is suitably modified. The modification is inspired by the recent results \cite{HLT24, BLRS25}.    
\end{itemize}

The $\multips_{\Lin'}$ lower bound result is obtained by combining the quadratic-size formula lower bound technique of Kalorkoti \cite{kal} with some additional ideas. The proof technique of $\ips_{\Lin'}$ lower bound result is inspired by the recent lower bound result of Chatterjee, Kush, Saraf and Shpilka \cite{CKSS24}.    
\end{abstract}

\newpage

\section{Introduction}
The main goal in propositional proof complexity is to prove lower bounds for computational resources required to prove propositional tautologies. 
This task in its full generality is strongly related to the separation problem of $\text{NP}$ and  $\text{coNP}$ as shown by Cook and Reckhow \cite{CR79}. 
Among the propositional proof systems, the Frege proof system is very well studied since the early work by Reckhow \cite{R76}. 
In Frege proofs, the propositions are computable by \emph{formulas} and lower bounds for the Frege system remain notoriously open. In the restricted setting, strong lower bounds for $\text{AC}^0$-Frege proof systems are known \cite{A88,BPI93,KPW95,PRST16,HR22,H23}.
In an attempt to address the Frege lower bounds via algebraic methods, Beame, Cook, and Hoover ~\cite{BCH1994}, and Pitassi and Impagliazzo ~\cite{PI1994}, introduced the Nullstellensatz proof system. 
The weak version of Hilbert's Nullstellensatz says that a set of polynomials (usually called \emph{axioms}) $f_1({\bar{x}}), \ldots, f_m(\bar{x})\in\mathbb{F}[x_1,\ldots, x_n]$ is unsatisfiable (over the algebraic closure of $\mathbb{F}$) if and only if there are polynomials $g_1(\bar{x}), \ldots, g_m(\bar{x})\in\mathbb{F}[x_1,\ldots, x_n]$ such that $\sum_{j=1}^m g_j(\bar{x}) f_j(\bar{x})=1$. 
The coefficients $g_1, \ldots, g_m$ are the Nullstellensatz refutations of the axioms. 
The \emph{degree} and the \emph{sparsity} are two important notions of the size measure for the refutations, and the lower bounds for them are known \cite{BIKPP94,BPRS97,R98,IPS99,BGIP01,AR01}. 
However the hard examples for such lower bound results usually admit polynomial-size Frege proofs. 
To overcome this, stronger algebraic proof systems were introduced that measure the size by the minimal size of the circuits computing the refutation polynomials $g_i(\bar{x})$. 
More precisely, this led to the Ideal Proof System (IPS) formulated by Grochow and Pitassi~\cite{GP18}, where the refutation of
$f_1({\bar{x}}), \ldots, f_m(\bar{x})$ over the Boolean hypercube satisfies the equation 
\[
\sum_i g_i(\bar{x}) f_i(\bar{x}) + \sum_j h_j(\bar{x})(x_j^2 - x_j) = 1,
\] 
and $g_i(\bar{x}), h_j(\bar{x})$ are represented by algebraic circuits. The size of an IPS refutation is the total size of the circuits computing the polynomials $g_i$ and $h_i$. 
Further formal description is given in \autoref{def:IPS}. 
The work of Grochow and Pitassi~\cite{GP18,Pit96} shows that $\ips$ is powerful enough to polynomially simulate the Frege and the Extended Frege systems, making $\ips$ lower bounds an important avenue of further research. The work of Forbes, Sphilka, Tzameret, and Wigderson~\cite{FSTW21} addresses fragments of linear $\ips$ in models like $\roabp$s, multilinear formulas (both unrestricted and constant depth), and for some classes of constant depth formulas. In another interesting line of work, Lee, Tzameret, and Wang~\cite{LTW18} connect the Frege lower bounds with noncommutative $\ips$ lower bounds.
More recently, the breakthrough result for the constant-depth circuit lower bounds by Limaye, Srinivasan, and Tavenas~\cite{LST21} led to a flurry of activities in this area~\cite{AF22,GHT22,HLT24}.
Even over fields of positive characteristic, new $\ips$ lower bounds were shown \cite{BLRS25,EGLT25}. These results were inspired by the result of Forbes~\cite{F24} on constant-depth circuit lower bounds in positive characteristic. 
It might be worthwhile to note, that none of these lower bounds are against polynomials that are arithmetizations of unsatisfiable CNFs. Therefore, it remains an important open problem to prove similar lower bounds for such polynomials.

In this paper, we address the $\ips$ lower bound problem in the setting of general algebraic formulas. Note that the current best-known (general) formula size lower bound for an explicit polynomial is given by techniques due to Kalorkoti~\cite{kal}\footnote{Also see \cite{CKSV22}, \cite[Section 3.2]{SY10}.}.
We lift this result to the $\ips$ setting by giving nearly similar quality lower bounds for general formulas, and the axiom polynomial is a variant of the subset-sum polynomial. More precisely, we prove the lower bound for $\multips_{\Lin'}$ as detailed in \autoref{subsec:results}. In fact, we are able to show such a lower bound (slightly weaker) even for a subset-sum axiom polynomial of \emph{constant degree}.  
In recent times, proving polynomial-quality lower bounds for constant-degree polynomials has received considerable attention since this is generally considered to be an avenue to improve the state-of-the-art polynomial-quality lower bounds for various algebraic models~\cite{HY09, CKV24}.  

Next, we consider the $\ips$ lower bound question for the sum-of-$\roabp$s model. 
Besides being a natural extension to the works that study the same question for $\roabp$s \cite{FSTW21, HLT24, BLRS25, EGLT25}, this model has an additional motivation in the context of $\VBP$ vs $\VNP$ conjecture. In particular, inspired by the connection that Bhargav, Dwivedi, and Saxena observe in the context of Valiant's $\VBP$ vs $\VNP$ conjecture \cite{BDS25}, the work of Chatterjee, Kush, Saraf, and Shpilka \cite{CKSS24} shows exponential-size lower bounds for sum-of-$\roabp$s model. 
Our result can be seen as a lift of those in \cite{CKSS24} to the $\ips$ setting. 

Both the results use the functional lower bound technique from \cite[Lemma 5.1]{FSTW21}. 
In particular, it shows that proving any refutation of the unsatisfiable system $\set{f=0,\  \bar{x}^2 - \bar{x}=0}$ is not in a circuit class $\ckt$ is equivalent to proving that each $g(\bar{x})$ which agrees with $\frac{1}{f(\bar{x})}$ over the Boolean hypercube, is not in $\ckt$. 

For our first result, to show that any multilinear-$\ips_{\Lin'}$ refutation of the system $\{f = 0,\, \bar{x}^2 - \bar{x} = 0\}$ requires a quadratic-size formula lower bound, note that it suffices to show that the unique multilinear polynomial $g(\bar{x})$ which agrees with $\frac{1}{f(\bar{x})}$ over the Boolean hypercube also requires a quadratic-size formula lower bound. 
This follows from the properties of the multilinear-$\ips_{\Lin'}$ model as explained in \autoref{def:IPS}, \autoref{def:IPS-restrict}. 
This was first formulated in the work of Govindasamy, Hakoniemi and Tzameret~\cite{GHT22}. 

For our second result, observe that the sum of $\roabp$s can be multilinearized, just like $\roabp$s (\autoref{prop:sum-roabp-mult-ub}). 
Hence, to show $\ips_{\Lin'}$ lower bound in the sum of $\roabp$s model for any refutation of the unsatisfiable system $\set{f=0,\ \bar{x}^2 - \bar{x}=0}$, it is sufficient to a lower bound on the size of any sum of $\roabp$s computing the unique multilinear polynomial $g(\bar{x})$ which agrees with $\frac{1}{f(\bar{x})}$ over the Boolean hypercube.  

\subsection{Our Results}\label{subsec:results}
We now state our main results.
\subsubsection*{Formula Lower Bounds }\label{sec:formula}
We first define the target polynomial $f$ for the unsatisfiable system. 
\begin{quote}
    Let $X = \set{x_1, x_2, \ldots, x_n}$ and $Y = \set{y_1, \ldots, y_{n-1}}$ be two sets of variables. 
    We partition $X$ into $N = \frac{n}{\log n}$ parts $\{X_1, \ldots, X_N\}$ where $X_i = \{x_{(i-1)\log n + 1}, \ldots, x_{i\log n}\}$ for each $i \in [N]$. 
    Clearly, the size of each $X_i$ is $\log n$.
    
    Fix any $i \in [N]$.
    For each subset $S \subseteq X_i$, let $\Pi_S \in \{0,1\}^{\log n}$ be its characteristic vector, and define $t(S) = \sum_{j=1}^{\log n} \Pi_S[j] \cdot 2^{j-1}$.
    Clearly, for any $s \in [0,\ldots,n-1]$ there is a unique $S$ such that $s = t(S)$.
    We define the polynomials,

    \begin{restatable}{restatableeqn}{eqpolytwo}
        \begin{equation*}
            f'(\vecx, \vecy) = \sum_{i=1}^N \sum_{S \subseteq X_i} \left( \prod_{j \in S} x_j \cdot y_{t(S)} \right) .
        \end{equation*}
    \end{restatable}
    \begin{restatable}{restatableeqn}{eqdefpolytwo}\label{eq:def-poly2}
        \begin{equation*}
        f(\vecx, \vecy)=\begin{cases}
            f'+1&\ \text{if the characteristic of }\ \mathbb{F}=0\\
            f'+\beta&\ \text{if the characteristic of}\ \mathbb{F}=p>0,\ k>0\in \mathbb{Z},\\ &\ \beta\in \mathbb{F}_{p^{k+1}}\setminus\mathbb{F}_{p^k}\ \text{and}\ f'\in \mathbb{F}_{p^k}[X,Y]
        \end{cases}
    \end{equation*}
    \end{restatable}     
\end{quote} 
Clearly, $f=0$ is unsatisfiable over the Boolean hypercube. 

Using this unsatisfiable system $\set{f=0,X^2 - X=0,Y^2 - Y=0}$, we show our first lower bound result.
Note that the degree of $f$ is $\log n + 1$ and sparsity is $O \inparen{\frac{n^2}{\log n}}$. 

\begin{restatable}{theorem}{formulaipslbhighdeg}\label{thm:formula-ips-lb-logn-deg}
    Consider the polynomial $f$ is defined in \autoref{eq:def-poly2}.
    Then the following statements are true :
    \begin{itemize}
        \item  Over the fields of characteristic $0$, any formula computing a $\multips_{\Lin'}$ refutation of the unsatisfiable system $\set{f=0,X^2 - X=0,Y^2 - Y=0}$ must have size at least $\Omega \inparen{\frac{n^2}{\log n}}.$   
        \item   Over the fields of characteristic $p$, any formula computing the $\multips_{\Lin'}$ refutation of the unsatisfiable system of equations $\set{f=0,X^2 - X=0,Y^2 - Y=0}$ must have size at least $\Omega\inparen{\frac{n^2}{\log n}}.$
    \end{itemize}
\end{restatable}

Next, we describe a constant degree set-multilinear polynomial which is unsatisfiable over the Boolean hypercube such that any multilinear proof of unsatisfiability requires near quadratic formula size. 
We first state a fact about the set-multilinear polynomials.

\begin{fact}\label{fact:total-sm-monomials}    
    Let $X = \bigsqcup_{i=1}^n X_i$ be a set of variables such that $|X_i|=\ell$ and $X_i=\setdef{x_{i,j}}{j\in [\ell]}$. 
    Then the number of set-multilinear monomials respecting the partition ${X_1, \ldots, X_n}$ is $\ell^n$. 
    Another way to see it is that any set-multilinear monomial $m = \prod_{i=1}^n x_{i,j_i}$ defines uniquely a map $\tau_m :[n]\rightarrow [\ell]$, with $\tau_m(i)=j_i$. 
    The number of such possible distinct maps is $\ell^n$. 
\end{fact}

We now define the constant-degree polynomial.

\begin{quote}
    Let $c>3$.
    Further, let $X = \setdef{x_{i,j}}{i \in [c], j\in [n^2/c]}$ and $Y = \setdef{y_{i,j}}{i,j\in [n]}$ be sets of variables. 
    The polynomial we will be defining, say $f \in \mathbb{F}[X,Y]$, is set-multilinear with respect to the partition $\set{X_1, \ldots, X_c, Y}$ where $X_i = \setdef{x_{i,j}}{j\in [n^2/c]}$.
    
    We further partition $X_i$ into $n^{2(1-1/c)}/c$ parts, each of size $n^{2/c}$, and denote it by $X_{i,k}$ for $k \in [n^{2(1-1/c)}/c]$. 
    So now, for any $i \in [c]$ and $k \in [n^{2(1-1/c)}/c]$, 
    \[
        X_{i,k} = \setdef{x_{i,j}}{j \in \set{(k-1)(n^{2/c}) + 1 , \ldots, k (n^{2/c})}}
    \]
    By \autoref{fact:total-sm-monomials}, it is easy to see that the number of set-multilinear monomials over the variable set $X^{(k)} = X_{1,k} \sqcup \cdots \sqcup X_{c,k}$ is $(n^{2/c})^c = n^2$. 
    
    Note that $|Y| = n^2$.
    Thus, for any $k \in [n^{2(1-1/c)}/c]$, we can define a bijection, say $\pi_k : \mathcal{M}_{\text{sm}}[X^{(k)}] \to Y$, which maps each set-multilinear monomial to a unique variable in $Y$. 
    For any such $k$, we define the set-multilinear polynomial $h_k$ of degree $c+1$, over the variable set $X_{1,k} \sqcup \cdots \sqcup X_{c,k} \sqcup Y$, as $h_k = \sum_{m \in \mathcal{M}_{\text{sm}}[X^{(k)} \sqcup Y]} m \cdot \pi_k(m)$.
    We then define the polynomial,
    \begin{restatable}{restatableeqn}{eqpolyone}\label{eq:poly1}
        \begin{equation*}
            h'(X,Y) = \sum_{k=1}^{n^{2(1-1/c)}/c} h_k . 
        \end{equation*}
    \end{restatable}
    \begin{restatable}{restatableeqn}{eqdefpolyone}\label{eq:def-poly1}
        \begin{equation*}
            h(X,Y)=\begin{cases}
                h'+1&\ \text{if the characteristic of }\ \mathbb{F}=0\\
                h'+\beta&\ \text{if the characteristic is}\ p>0,\ k>0\in \mathbb{Z},\\ &\ \beta\in \mathbb{F}_{p^{k+1}}\setminus\mathbb{F}_{p^k}\ \text{and}\ h'\in \mathbb{F}_{p^k}[X,Y]
            \end{cases}
        \end{equation*}
    \end{restatable}
\end{quote}

Note that the system of equations $\set{h=0,X^2 - X=0,Y^2 - Y=0}$ is clearly unsatisfiable. 
Further $h$ is a polynomial over $n^2$ variables of sparsity $n^{4 - 2/c}/c$ and degree $c+1$. 
Using this system, we show our second lower bound result.
\begin{restatable}{theorem}{formulaipslbconstantdeg}\label{thm:formula-ips-lb-const-deg}
    Consider the polynomial $h$ in \autoref{eq:def-poly1}. Then the following statements are true.
    \begin{itemize}
        \item  Over the fields of characteristic $0$, any formula computing a $\multips_{\Lin'}$ refutation of the unsatisfiable system $\set{h=0,X^2 - X=0,Y^2 - Y=0}$ must have size at least $\Omega(n^{4-2/c}).$
        \item Over the fields of characteristic $p$, any formula computing a $\multips_{\Lin'}$ refutation of the unsatisfiable system $\set{h=0,X^2 - X=0,Y^2 - Y=0}$ must have size at-least $\Omega\inparen{n^{4-2/c}}$.
    \end{itemize}
\end{restatable}

\begin{remark}\label{rmk:0}
Since $f$ and $h$ are sparse polynomials, they can be obtained by substitutions of monomials in the usual subset-sum axiom polynomial of the form $\sum_{i} z_i -\gamma$. See \autoref{sec:IPS-UB} for further details. 
\end{remark}
\begin{remark}\label{rmk:1}
We also note that the unsatisfiable systems given by $f$ and $h$ have \emph{non-multilinear} refutations of polynomial size constant depth formulas. This follows easily from the known results \cite{FSTW21,BLRS25}. 
A proof sketch is given in \autoref{sec:IPS-UB} for completeness.
\end{remark}

\subsubsection*{Sum of ROABPs Lower Bounds}\label{sec:sum-roabp}
Over the fields of characteristic $0$, we use a variant of the subset-sum polynomial to show exponential-size lower bound for the $\ips_{\Lin'}$ refutations in the sum of $\roabp$s model. 
The same polynomial has been used earlier to prove exponential-size lower bound for $\ips_{\Lin'}$ refutations in (any order) $\roabp$s model \cite{FSTW21}.
More precisely, we prove the following.

\begin{restatable}{theorem}{sumofroabp}\label{thm:sum-roabp-ips-lb}
    For $X = \set{x_0, \ldots, x_{2n-1}}$, $T = \setdef{t_{i,j}}{i, j \in [0, \ldots, 2n-1] \text{ with } i< j}$ and $\beta = 2 {2n \choose 2}$, let $f \in \mathbb{F}[X,T]$ be the polynomial defined as 
    \[
        f = \inparen{\sum_{0 \leq i < j \leq 2n-1} t_{i,j} x_ix_j} - \beta
    \]
    which is unsatisfiable over the Boolean hypercube. 
    Then there exists $\gamma>0$ such that the total width of any sum of $\roabp$ computing the linear proof of unsatisfiability $(\ips_{\Lin'})$, for the system $\{f=0,X^2 - X=0,T^2-T=0\}$, is at-least $\exp(n^{\gamma})$.
\end{restatable}

Similar lower bounds can be achieved over the fields of positive characteristic as well. 
We refer the readers to \autoref{sec:sum-of roabp-ips-lb-positive-char} for further details.
\begin{remark}\label{rmk:2}
Further, we note that there is a \emph{non-multilinear} refutation of the unsatisfiable system given by $f$ which is computable by  $\poly(n)$ size ABP. 
More details can be found in \autoref{sec:IPS-UB}. 
\end{remark}

\subsection{Proof Sketches}
\paragraph*{Formula lower bounds} The main ideas behind the $\multips_{\Lin'}$ lower bounds in the formula model (\autoref{thm:formula-ips-lb-logn-deg}, and \autoref{thm:formula-ips-lb-const-deg}) are based on the techniques from \cite{kal} and the functional lower bound method of \cite{FSTW21}. The functional method shows that, the proof of a $\ckt\text{-}\ips_{\Lin'}$ lower bound for the unsatisfiable system  $\set{f=0,x_1^2-x_1=0,\ldots,x_n^2-x_n=0}$, is equivalent to the fact that every polynomial $g(\bar{x})$ that agrees with $\frac{1}{f(\bar{x})}$ over the Boolean hypercube satisfies that $g\notin\ckt$. Since our goal is to show a formula lower bound in the $\multips_{\Lin'}$ model, it suffices to show that the unique multilinear polynomial $g$ that agrees with $\frac{1}{f}$ over the Boolean hypercube requires nearly quadratic size formulas.

Let $X=\set{x_1,\ldots,x_n}$ and $f\in \mathbb{F}[X]$. 
For a subset $S\subseteq X$, one can express the polynomial $f = \sum_{m\in \mathcal{M}[S]} m\cdot f_m$, where $\mathcal{M}[S]$ is the set of monomials defined on $S$ and $f_m\in \mathbb{F}[X\setminus S]$. 
Let $\coeff_S(f) \eqdef \set{f_m\ :\ m\in \mathcal{M}[S],\ f_m \neq 0}$ and $\algrank_S(f)$ be the algebraic rank of $\coeff_S(f)$. 
The main result of \cite{kal} states that, if $f$ is computable by a formula of size $s$, then for any partition $X_1\sqcup X_2\sqcup\ldots\sqcup X_t$, we have that $s\geq \Omega\inparen{ \sum_{i=1}^t \algrank_{X_i}(f)}$.

It is a well-known fact that for any set of polynomials, the algebraic rank is lower bounded by the algebraic rank of their leading or trailing monomials. We would like to apply Kalorkoti's method on the unique multilinear polynomial $g$ that agrees with $\frac{1}{f}$ over the Boolean hypercube. To do that, we need to study the relationship between the support structure of $f$ and $g$.

Note that our polynomials have the property that for any two monomials, the support of one is not contained in the other. 
We show that for any polynomial $f$ that satisfies this property, $\supp(f)\subseteq \supp(g)$  (\autoref{obs:support-obs-1}). 
Furthermore, if a monomial $m$ cannot be written as a multilinearized product of monomials from $\supp(f)$, then its coefficient in $g$ is $0$ (\autoref{lem:support-lem-1}).
Using these, it can be observed that for any $S\subseteq X$, the trailing monomials (under graded-lex ordering) of $\coeff_S(g)$ are equal to the trailing monomials of $\coeff_S(f)$. 

Under the partition $X_1\sqcup\cdots\sqcup X_N$ considered in the definition of \autoref{eq:def-poly2}, since $\coeff_{X_i}(f)$  is a set of algebraically independent monomials, the trailing monomials in $\coeff_{X_i}(g)$ are also algebraically independent. The result now follows from \cite{kal}.

\paragraph*{Sum of ROABPs Lower Bounds} The $\ips_{\Lin'}$ lower bound for the sum of $\roabp$s model combines techniques from the works of Forbes, Shpilka, Tzameret and Wigderson~\cite{FSTW21} and Chatterjee, Kush, Saraf, and Shpilka~\cite{CKSS24}. Similar to the formula setting, it is enough to show a lower bound against all polynomials $g$ that agree with $1/f$ over the Boolean hypercube. Additionally, since sum of $\roabp$s can be efficiently multilinearized (\autoref{prop:sum-roabp-mult-ub}), it is enough to show a lower bound against the unique multilinear polynomial satisfying the above property. 

The proof broadly consists of two parts. The first part is to establish a structural weakness of the sum of $\roabp$s. Roughly speaking, under a random partition of the variables, the rank of the partial derivative matrix of a polynomial computed efficiently by a sum of $\roabp$s is low with high probability (\autoref{lem:weakness-sum-roabp}). The proof is developed using the ideas implicit in \cite{CKSS24}. On the other hand, for any balanced partition, the derivative matrix corresponding to the unique multilinear polynomial $g$ that agrees with $1/f$ (where $f$ is defined in \autoref{thm:sum-roabp-ips-lb}) has high rank (\cite{FSTW21}). The lower bound follows by combining these two statements. 

\subsection*{Organization}
In \autoref{sec:prelim}, we present the necessary definitions and preliminary concepts. \autoref{sec:formula-ips-lin-lb} details the proofs of our results on formula lower bounds. 
The results on the sums of ROABPs is presented in \autoref{sec:sum-roabp-ips-lin-lb}. 
The conclusion raises a few questions for further study and the appendix provides some additional observations.

\section{Preliminaries}\label{sec:prelim}

We begin by stating the Chernoff Bound.
\begin{theorem}[Chernoff Bound]\label{thm:chernoff-bound}
    Let $X_1,\ldots,X_n$ be a set of $0$-$1$ independent random variables. Let $X=\sum_{i=1}^n X_i$ and $\mu=\mathbb{E}[X]$ is the expected value of $X$. Then for any $\delta \in (0,1)$,
    \[
        \Pr[X\leq (1-\delta)\mu]\leq \exp\inparen{-\frac{\delta^2\mu}{2}}.
    \]
\end{theorem}

\subsection{Notations}\label{sec:notations}
For a set $X$, a partition is written as $X = X_1 \sqcup \cdots \sqcup X_k$. The set of natural numbers is $\mathbb{N}$. For $n \in \mathbb{N}$, the set $\set{1, \ldots, n}$ is denoted by $[n]$. 
    The symmetric group over $\{1,2,\ldots,n\}$ is $S_n$. For notational clarity, sometime $\bar{x}$ or $X$ is used for the set of variables $\{x_1,\ldots,x_n\}$.  
    The set of all possible monomials over $X$ is denoted by $\mathcal{M}(X)$. The set of equations $\setdef{x^2_i-x_i=0}{i\in [n]}$ is sometime shorten as  $X^2 - X$. For $\set{a_1, \ldots, a_n} \subseteq \mathbb{N}$, we denote by $\vecx^{\veca}$, the monomial $\prod_{i \in [n]} x_i^{a_i}$.
    Given a monomial $\vecx^{\veca}$,  define $\supp(\vecx^{\veca})$ to be the set $\setdef{x_i}{a_i \geq 1}$. Given a monomial $\vecx^{\veca}$, we define $\mult(\vecx^{\veca})$ to be the multilinear version of the monomial.
    That is, $\mult(\vecx^{\veca}) = \prod_{i=1}^n x_i^{\min\{a_i,1\}}$.
    By linearity, we extend this to define $\mult(f)$ for any polynomial $f \in \mathbb{F}[X]$. 
    
    For a subset $S\subseteq [n]$, we denote by $\mathbb{1}_S: [n] \to \set{0,1}$ the characteristic function for $S$. That is, for any $i \in [n]$, $\mathbb{1}_S(i) = 1 \iff i \in S$.

\subsection{Definitions}\label{sec:definitions}
\subsubsection*{Models of Computation}

\begin{definition}[Algebraic Formulas]
\label{def:formula}
An algebraic formula $C$ is a directed tree with a unique output gate (root) of out-degree $0$, and input gates of in-degree $0$ (leaves) labeled by variables $x_1, \ldots, x_n$ or constants from $\mathbb{F}$. The internal gates are labeled by $+$ or $\times$. Each gate $v$ computes a polynomial $f_v$ defined recursively: if $v$ is an input, then $f_v = \textsf{label}(v) \in \{x_1, \ldots, x_n\} \cup \mathbb{F}$; if $v = u\ \textsf{op}\ w$ for $\textsf{op} \in \{+, \times\}$, then $f_v = f_u\ \textsf{op}\ f_w$. The polynomial computed at the output gate is the polynomial computed by the formula.
\end{definition}

\begin{definition}[ROABPs]\label{def:roabp}
    A Read-once Oblivious ABP \emph{($\roabp$)} is a directed acyclic graph where the vertex set is partitioned into layers $0, 1, \ldots, n$ with directed edges only between adjacent layers ($i$ to $i+1$).
    Layers $0$ and $n$ have a single vertex each (called the source $s$ and terminal $t$ respectively), whereas the other layers can have any number of vertices.
    
    The labels on the edges satisfy the property that for every $i \in [n]$, there is a unique $j \in [n]$ such that every edge in between layer $j-1$ and $j$ is labelled by a univariate polynomial in $x_i$.
    The polynomial computed by any $s$-to-$t$ path is the product of the edge labels on it and the polynomial computed by the $\roabp$ is the sum of all polynomials computed by such paths.
    
    An $\roabp$ is said to have order $\sigma \in S_n$ if for every $j \in [n]$, the edges in between layers $j-1$ and $j$ are labelled by univariates in $x_{\sigma(j)}$. 
    An $\roabp$ is said to be multilinear if each of the edge labels are linear polynomials. 

    The size of an $\roabp$ is the total number of vertices in it; the width of any layer in the $\roabp$ is the number of vertices in it and the width of an $\roabp$ is the width of its widest layer.   
    A sum of $\roabp$s is defined as the sum of individual $\roabp$s, with total width equal to the sum of their widths.
\end{definition}

\begin{definition}[Set-Multilinear Polynomial]\label{def:setmultilinear}
   Let $(X_1, \ldots, X_d)$ be a partition of the variable set $X$, with $X_i=\{x_{i,1}, x_{i,2}, \ldots, x_{i,n}\}$. 
   A polynomial $f\in\mathbb{F}[X]$ is said to be set-multilinear with respect to the given partition if each monomial in $f$ is of the form $(x_{1, j_1} x_{2, j_2}\cdots x_{d, j_d})$.  
\end{definition}

It is sometimes useful to think of the variables as coming from a matrix $M_{d \times n}$ where the $i^{th}$ row is $\set{x_{i,1}, x_{i,2}, \ldots, x_{i,n}}$ and a set-multilinear polynomial is one in which each monomial is constructed by picking exactly one variable from each row.

\subsubsection*{Ideal Proof System}

We begin with the definition of the Ideal Proof System $(\ips)$ and some of its restrictions.

\begin{definition}[Ideal Proof System \cite{FSTW21,GP18}]\label{def:IPS} 
    Let $f_1, \ldots, f_m\in \mathbb{F}[X]$ be a set of polynomials such that $\set{f_1, \ldots, f_m, x_1^2 - x_1, \ldots, x_n^2 - x_n}$ has no common solution over the Boolean hypercube\footnote{That is, there does not exist $\vecx \in \{0,1\}^n$ such that for every $i \in [m]$, $f_i(\vecx) = 0$.}.
 
    A \emph{proof} of the unsatisfiability of this set of polynomial equations, in the \emph{Ideal Proof System} $(\ips)$, is a polynomial $P(X, y_1, \ldots, y_m, z_1, \ldots, z_n) \in \mathbb{F}[X,Y,Z]$ such that the following holds:
     \begin{itemize}\itemsep0em
        \item $P(X,\bar{0},\bar{0})=0$;
        \item $P(X, f_1, \ldots, f_m, x_1^2-x_1, \ldots, x_n^2-x_n) = 1$. \qedhere
     \end{itemize}
\end{definition}

\begin{definition}[Restrictions of IPS \cite{GHT22,HLT24}]\label{def:IPS-restrict}
    Some restrictions of the Ideal Proof System that we will be considering are as follows.
    \begin{itemize}\itemsep0em
        \item $\ips_{\Lin}$: A proof, $P$, in the Ideal Proof System is said to be in $\ips_{\Lin}$ if it additionally satisfies the conditions $\forall i \in [m],\ j\in [n],\  \deg_{y_i}(P), \deg_{z_j}(P) \leq 1$.
        \item $\ips_{\Lin'}$: A proof, $P$, in the Ideal Proof System is said to be in $\ips_{\Lin'}$ if it only satisfies the conditions $\forall i \in [m], \deg_{y_i}(P) \leq 1$.
        \item $\multips_{\Lin'}$: A proof, $P$, in $\ips_{\Lin'}$ is said to be in $\multips_{\Lin'}$ if it additionally satisfies the condition that $P(X,Y,\bar{0})$ is multilinear polynomial.
        Note that $P(X,\bar{0},Z)$ is not necessarily multilinear.
        \item $\ckt$-$\ips_{\Lin'}$ and $\ckt$-$\multips_{\Lin'}$: For any polynomial class $\ckt$, a proof in $\ips_{\Lin'}$ is said to be in $\ckt$-$\ips_{\Lin'}$ if it is additionally contained in $\ckt$. 
        $\ckt$-$\multips_{\Lin'}$ is defined analogously. \qedhere
    \end{itemize}
\end{definition} 

\subsubsection*{Monomial Ordering}

\begin{definition}[Monomial Ordering]\label{def:mon-order}
    Given a set of variables $X = \set{x_1,\ldots,x_n}$, let $(x_{i_1}, \ldots, x_{i_n})$ be a total ordering on it.
    We extend it to a total ordering $\succ$ on $\mathcal{M}(X)$ as follows.
    \begin{quote}
        For any two distinct monomials $\vecx^{\veca}, \vecx^{\vecb}$, 
        \begin{itemize}\itemsep0pt
            \item if $\deg(\vecx^{\veca}) > \deg(\vecx^{\vecb})$ then $\vecx^{\veca} \succ \vecx^{\vecb}$;
            \item if $\deg(\vecx^{\veca}) = \deg(\vecx^{\vecb})$, $a_{i_j} = b_{i_j}$ for every $j < j_0$ and $a_{i_{j_0}} > b_{i_{j_0}}$, then $\vecx^{\veca} \succ \vecx^{\vecb}$.
        \end{itemize}
    \end{quote}
    For a polynomial $f \in \mathbb{F}[X]$ and a total order $\succ$ on $\mathcal{M}[X]$, we denote the \emph{leading monomial} of $f$ (largest monomial under $\succ$ that present in $f$) and \emph{trailing monomial} of $f$ (smallest monomial under $\succ$ that is present in $f$) by $\LM(f)$ and $\TL(f)$ respectively.
    
    For a subset $S\subseteq \mathbb{F}[{X}]$, we define $\LM(S):= \setdef{\LM(f)}{f \in S}$.
    $\TL(S)$ is defined similarly.
\end{definition}

\subsubsection*{Algebraic Independence and Algebraic Rank}
\begin{definition}\label{def:alg-ind} 
    A set of polynomials $\set{f_1, \ldots, f_m} \subseteq \mathbb{F}[X]$ is said to be algebraically dependent over $\mathbb{F}$ if there exists a non-zero polynomial $A(y_1,\ldots,y_m) \in \mathbb{F}[Y]$ such that $A(f_1,\ldots,f_m) = 0$.
    If no such $A$ exists, then $f_1, \ldots, f_m$ are said to be algebraically independent.
    
    Given any set of polynomials $S \subseteq \mathbb{F}[X]$, the algebraic rank of $S$ is the size of largest algebraic independent subset of $S$. 
\end{definition}

\begin{definition}\label{def:tr-deg} 
    Given a polynomial $f \in \mathbb{F}[X]$ and $S \subseteq X$, let $\mathcal{M}[S]$ be the set of all monomials that can be defined over the variables in $S$.
    Furthermore, for each $m \in \mathcal{M}[S]$, let $f_m \in \mathbb{F}[X \setminus S]$ be the unique polynomial such that $f = \sum_{m\in \mathcal{M}[S]} m \cdot f_m$.
    We define $\algrank_S(f)$ as the algebraic rank of the set $\coeff_S(f) \eqdef \setdef{f_m}{m \in \mathcal{M}[S],\ f_m \neq 0}$. 
\end{definition}
Here is a standard theorem on algebraic independence, that we need.
 \begin{theorem}[\cite{KR05}]
     \label{thm:alg-ind-LM,TM} Let $f_1,\ldots,f_k\in \mathbb{F}[X]$ and $\succ$ be a total ordering over the monomials. If $f_1,\ldots,f_k$ are algebraically dependent then both sets $\{\LM(f_i)\ |\ i\in [1,2,\ldots,k]\}$ and $\{\TL(f_i)\ |\ i\in [1,2,\ldots,k]\}$ are algebraically dependent. 
 \end{theorem}

\subsection*{Partial Derivative Matrix}

\begin{definition}[Partial Derivative Matrix \cite{SY10,rpsurvey}]\label{def:PD-matrix} 
    Let $f\in \mathbb{F}[X]$ be a polynomial and $(Y,Z)$ be a partition of $X$ (that is, $X = Y \sqcup Z$). 
    The partial derivative matrix of $f$ with respect to $(Y, Z)$, say $M_{Y,Z}(f)$, is defined as follows.
    \begin{itemize}\itemsep0pt
        \item The rows of $M_{Y,Z}(f)$ are indexed by monomials in the variables $Y$ and the columns are indexed by monomials in the variables $Z$. 
        \item Given monomials $m_Y = \vecy^{\veca}$ and $m_Z=\vecz^{\vecb}$, the entry of $M_{Y,Z}(f)$ in the row labelled $m_Y$ and column labelled $m_z$ is the coefficient of $m_Y \cdot m_Z$ in $f$ (denoted by $\coeff_{m_Y \cdot m_Z}(f)$). \qedhere
    \end{itemize} 
\end{definition}

Given a partition $(Y,Z)$ of $X$ and $f \in \mathbb{F}[X]$, note that we can also consider $f$ to be a polynomial over the variables $Z$ with coefficients being polynomials over in the variables $Y$ variables.
We define $\coeff_{Y,Z}(f) = \setdef{\sum_{m_Y \in \mathcal{M}(Y)} m_Y \cdot \coeff_{m_Y \cdot \vecz^{\vecb}}(f)}{\vecz^{\vecb} \in \mathcal{M}(Z)}$. 
\begin{definition}[Evaluation Dimension\cite{FSTW21}]
\label{def:eval-dim}
Let $f(X,Y)\in \mathbb{F}[X,Y]$ be a polynomial and $S\subseteq \mathbb{F}$. We define evaluation dimension of $f$ on partition $(X,Y)$ in the following way,
\[
\eval_{X,Y,S}[f(\vecx,\vecy)]=\dim_{\mathbb{F}}\set{f(\vecx,\bar{\beta}): \bar{\beta}\in S^{|Y|}} \qedhere
\]
\end{definition}

\begin{lemma}[\cite{FS15}]\label{lem:rank-eval-dim-lb}
Let $f\in \mathbb{F}[X,Y]$ and $S$ be any subset of $\mathbb{F}$. Then 
\[
\dim_{\mathbb{F}}\inparen{\coeff_{X,Y}(f)}\geq \eval_{X,Y,S}(f).
\] 

\end{lemma}

\subsection{Functional Lower Bound Method for proving lower bounds for IPS proofs}

We will be crucially using the following technique described in \cite{FSTW21} for proving lower bounds against $\ips_{\Lin}$ and $\ips_{\Lin'}$.

\begin{restatable}{theorem}{funclbmethod}\emph{(\cite[Lemma 5.1]{FSTW21})}\label{thm:func-lb-ips-connection}
    Let $f\in \mathbb{F}[X]$ be a polynomial such that for some $\beta\neq 0\in \mathbb{F} $ the system $\set{f - \beta, X^2 - X}$ has no common solutions over the Boolean hypercube. 
    Further, let $\ckt \subseteq \mathbb{F}[X]$ be a class of polynomials that is closed under partial $\mathbb{F}$-assignments\footnote{If $f \in \mathbb{F}[X]$ with $f \in \ckt$ and $Y \subseteq X$, then $f|_{Y = \veca}(X \setminus Y) \in \ckt$ for any $\veca \in \mathbb{F}^{\abs{Y}}$.}. 
    
    If there does not exist any $g \in \ckt$ that satisfies $g(\vecx) = \frac{1}{f(\vecx) - \beta}$ for every $\vecx \in \{0,1\}^n$, then there is no proof of unsatisfiability for the system $\set{f - \beta, X^2 - X}$ that is contained in $\mathcal{C}\text{-}\ips_{\Lin}$, $\mathcal{C}\text{-}\ips_{\Lin'}$.
\end{restatable}
This is called the functional lower bound method, since one needs to prove a lower bound against all polynomials which evaluate to the same value as $\frac{1}{f-\beta}$ over the entire Boolean hypercube. 
\begin{proposition}\label{prop:mult-ips-functional-lb}
    Let $f\in \mathbb{F}[X]$ and $\set{f=0,X^2 - X=0}$ be a unsatisfiable system against which we want to show $\ckt\text{-}\multips_{\Lin'}$ lower bound. Let $g(X)$ be the unique multilinear polynomial that agrees with $\frac{1}{f(X)}$ over the Boolean hypercube. If $g\notin \ckt$ then the system does not have $\ckt\text{-}\multips_{\Lin'}$ refutation. 
\end{proposition}

\begin{proof}
    Let $C(X,y,Z)$ be a $\ckt\text{-}\multips_{\Lin'}$ refutation for the unsatisfiable system. Assume $g(X)$ be the unique multilinear polynomial such that $g(X)=\frac{1}{f(X)}\mod{X^2 - X}.$ Since, $f$ is the only non-Boolean axiom and $C$ is linear in $y$, by \autoref{def:IPS-restrict}, $C(X,y,Z)=g(X)\cdot y+ C'(X,Z,y)$ and $C(X,f,X^2 - X)= g(X)\cdot f(X)+\sum_{i=1}^n C'_i(X)(x_i^2-x_i)=1.$ So, $C(X,y,0)=y\cdot g(X)\implies C(X,1,0)=g(X).$ This implies $g(X)\in \ckt$, but this contradicts the assumption : $g\notin \ckt.$
\end{proof}


\section{Lower Bound Against Formulas for $\multips_{\Lin'}$ Proofs}\label{sec:formula-ips-lin-lb}
In this section, we prove \autoref{thm:formula-ips-lb-logn-deg} and \autoref{thm:formula-ips-lb-const-deg}. 

Given an unsatisfiable system $\set{f=0,X^2 - X=0}$, to prove the formula complexity lower bound of the $\multips_{\Lin'}$ refutation for $f$, it suffices to consider the unique multilinear polynomial $g(X)$ such that, 
$$g(X)=\frac{1}{f(X)}\mod{X^2 - X}.$$ Towards that we first note a few structural properties of the polynomial $g(X)$.
\subsection{Some Structural Results}\label{subsec: struct-result}

\begin{proposition}\label{obs:support-obs-1}
    Let $\mathbb{F}$ be any field and $f\in \mathbb{F}[{X}]$ be a multilinear polynomial such that the system of equations $\{f=0,X^2 - X=0\}$ is unsatisfiable over $\set{0,1}^n$. 
    Additionally, suppose $f$ has the property that for any two monomials $m_i,m_j\in f$ with non empty support, $\supp(m_i)\nsubseteq \supp(m_j)$ and $\supp(m_j)\nsubseteq\supp(m_i)$. 
    Let $g\in \mathbb{F}[{X}]$ be the unique multilinear polynomial such that $g(X)= \frac{1}{f(X)}\mod{X^2 - X}$.

    Then, for any monomial $m$ whose coefficient in $f$ is non-zero, its coefficient in $g$ is also non-zero\footnote{If $f$ is not multilinear, then for any non-multilinear monomial $m$ with non-zero coefficient in $f$, the coefficient of $\mult(m)$ in $g$ is non-zero. }.
\end{proposition}
 
\begin{proof}
    Let $f(X)=f'(X)-\beta$ where $f(0)=-\beta$. 
    Note that since $f$ is unsatisfiable, $\beta\neq 0$. The unique multilinear polynomial $g(X)=\frac{1}{f(X)}\mod{X^2 - X}$ is given by the following.
    \[
        g({\vecx})= \sum_{T\subseteq [n]} g(\mathbb{1}_T)\prod_{i\in T}x_i\prod_{i\notin T}(1-x_i).
    \]

    Let $m$ be a monomial with non-zero coefficient $\alpha$ in $f$ and $S=\supp(m)$. 
    Setting the variables $x_i\notin S$ to $0$, it is easy to see that the coefficient of $m$ in $g$ (denoted by $c_m$), is given by the expression $c_m=\sum_{A\subseteq S}g(\mathbb{1}_A)(-1)^{|S\backslash A|}$.
    Notice that in the expression $c_m$, the term in the summand is $g(\mathbb{1}_S)=\frac{1}{f(\mathbb{1}_S)}=\frac{1}{\alpha-\beta}$ when $A=S$ and it is $g(\mathbb{1}_\emptyset)(-1)^{|S|}=\frac{(-1)^{|S|}}{-\beta}$ when $A=\emptyset$. 
    Moreover, from the monomial support property, for any $S'\subset S$, $g(\mathbb{1}_{S'})=-\frac{1}{\beta}$. 
    Thus,
    \begin{equation}\label{eq:supp-obs-1}
        \begin{split}
            c_m=\sum_{A\subseteq S}g(\mathbb{1}_A)(-1)^{|S\backslash A|}= \frac{1}{\alpha-\beta} -\frac{1}{\beta}\sum_{A\subset S}(-1)^{|S\backslash A|}\\= \frac{1}{\alpha-\beta}-\frac{1}{\beta}\inparen{\sum_{A\subseteq S}(-1)^{|S\setminus A|}-1}=\frac{1}{\alpha-\beta}+\frac{1}{\beta}.
        \end{split}
    \end{equation}
    In the above \autoref{eq:supp-obs-1}, we use the standard fact that 
    \[
        \sum_{A\subseteq S} (-1)^{|S\setminus A|}=\sum_{i=0}^{|S|}\binom{|S|}{i}(-1)^{|S|-i}=0. \qedhere
    \]
\end{proof}

We record some further structural properties in the following lemma. The lemma shows that any monomial $m$ which is not expressible by the multilinearization of any product of monomials from $\supp(f)$, has coefficient $0$ in $g$. 
 \begin{lemma}\label{lem:support-lem-1}
          Let $\mathbb{F}$ be any field and $f\in \mathbb{F}[{X}]$ be a multilinear polynomial such that $\{ f=0,X^2 - X=0\}$ is unsatisfiable. Let $g\in \mathbb{F}[{X}]$ be the unique multilinear polynomial such that $g(X)= \frac{1}{f(X)}\mod{X^2 - X}$. Consider a monomial $m$ such that $m\neq\mult(\prod_{m\in S}m)$ for any subset $S\subseteq \supp(f)$. Then the coefficient of $m$ in $g$ is $0$. 
 \end{lemma}
\begin{proof}
    Consider a monomial $m$ such that $m\neq\mult(\prod_{m\in S}m)$ for any subset $S\subseteq \supp(f)$. 
    The idea is to decompose $m$ as a product of monomials $m_1\ \text{and}\ m_2$ with the following property : 
    There is a set $S_1\subseteq\supp(f),\ \text{such that}\  m_1=\mult(\prod_{m'\in S_1} m')$ and $S_2=\supp(m_2)\notin \supp(f)$. 
    Moreover for any subset $T_1\subseteq S_1$ and a nonempty subset $ T_2\subseteq S_2$, $T_1\cup T_2\notin \supp(f)$. 
    It is not hard to see that such a decomposition can be constructed in a greedy manner.
   
    Next, we make a few simple observations. Clearly, $S_1\cap S_2=\emptyset$ due to multilinearity. Moreover, $g(\mathbb{1}_{S_2})= 1/f(0)$. 
    Furthermore, for any subset $T_1\subseteq S_1$ and $T_2\subseteq S_2$, 
    \[
        g(\mathbb{1}_{T_1\cup T_2})=\frac{1}{f(\mathbb{1}_{T_1\cup T_2})}=\frac{1}{f(\mathbb{1}_{T_1})}.
    \]
    Recall that, $$g(\vecx)= \sum_{T\subseteq [n]} g(\mathbb{1}_T)\prod_{i\in T}x_i\prod_{i\notin T}(1-x_i).$$ 
    Hence, 
      \begin{align*} 
         c_m=\sum_{A\subseteq S_1\cup S_2}g(\mathbb{1}_A)(-1)^{|S_1\cup S_2\backslash A|}
         &= \sum_{A_1\subseteq S_1}(-1)^{|S_1\backslash A_1|}\sum_{A_2\subseteq S_2}g(\mathbb{1}_{A_1\cup A_2})(-1)^{|S_2\backslash A_2|}\\
         &=\sum_{A_1\subseteq S_1}(-1)^{|S_1\backslash A_1|}\sum_{A_2\subseteq S_2}g(\mathbb{1}_{A_1})(-1)^{|S_2\backslash A_2|}\\
         &=\sum_{A_1\subseteq S_1}g(\mathbb{1}_{A_1})(-1)^{|S_1\backslash A_1|}\sum_{A_2\subseteq S_2}(-1)^{|S_2\backslash A_2|}=0.
    \end{align*}
    Here we have used the fact that $\sum_{A_2\subseteq S_2}(-1)^{|S_2\backslash A_2|}=0$ and $g(\mathbb{1}_{A_1\cup A_2})=g(\mathbb{1}_{A_1})$.
\end{proof}


 \subsection{The Lower Bound}
 In this section, we prove a near-quadratic $\multips_{\Lin'}$ size lower bound in the formula setting. We use the structural results developed in \autoref{subsec: struct-result} along with the lower bound technique of \cite{kal}.
\begin{theorem}[Kalorkoti, \cite{kal}]
   \label{thm:kalorkoti}
   Let $f\in \mathbb{F}[X]$ be a polynomial computed by a size $s$ formula. 
   Let $(X_1, X_2, \ldots, X_t)$ be any partition of the variables set $X$. 
   Then $s$ is at-least $\Omega\inparen{ \sum_{i=1}^t \algrank_{X_i}(f)}$.
\end{theorem}
Next, we recall the axiom polynomial from \autoref{subsec:results}, for which we show the lower bound result for $\ips_\Lin'$ refutations.

Let $X = \set{x_1, x_2, \ldots, x_n}$ and $Y = \set{y_0, \ldots, y_{n-1}}$ be two sets of variables. 
We partition $X$ into $N = \frac{n}{\log n}$ parts, $\{X_1, \ldots, X_N\}$ where $X_i = \{x_{(i-1)\log n + 1}, \ldots, x_{i\log n}\}$. 
Clearly, the size of each $X_i$ is $\log n$.
    
For a subset $S \subseteq X_i$, let $\Pi_S \in \{0,1\}^{\log n}$ be its characteristic vector, and define $t(S) = \sum_{j=1}^{\log n} \Pi_S[j] \cdot 2^{j-1}$.
Now we define the polynomial, 
\eqpolytwo* 
\eqdefpolytwo*

We now prove the first lower bound of this section: a $\log n$-degree unsatisfiable system whose $\multips_{\Lin'}$ refutation by formulas requires near-quadratic size. We first recall the statement.
\formulaipslbhighdeg*
\begin{proof}
    Let $g(X,Y)$ be the unique multilinear polynomial such that $g(X,Y) = \frac{1}{f(X,Y)} \mod {X^2 - X, Y^2 - Y}$. For any two monomials $m_1,m_2\in \supp(f)$, $\supp(m_1)\nsubseteq\supp(m_2)\ \text{and}\ \supp(m_2)\nsubseteq\supp(m_1)$. 
    Hence \autoref{obs:support-obs-1} implies that, all monomials of $f$ appear in $g$ with non-zero coefficients. Moreover, \autoref{lem:support-lem-1} implies every other monomial in $g$ with non-zero coefficient can only be the multilinearized product of monomials from $\supp(f)$.
    Consider the variable partition $X = X_1 \sqcup X_2 \sqcup \cdots \sqcup X_N$ as defined above, and fix any $X_i$. 
    Order monomials with $X \succ Y$, then extend as in~\autoref{def:mon-order}, using any order within $X$ and $Y$. 
    Under this ordering, $\TL(\coeff_{X_i}(g)) = \set{y_0, \ldots, y_{n-1}}$, which is algebraically independent. 
    So using \autoref{thm:alg-ind-LM,TM}, the set $\coeff_{X_i}(g)$ is also algebraically independent. 
    Let $g$ be computable by a formula of size $s$. Then using \autoref{thm:kalorkoti}, we conclude 
    \[
        s\geq \Omega\inparen{\sum_{i=1}^N \algrank_{X_i}(g)+\algrank_Y(g)}=\Omega\inparen{\frac{n^2}{\log n}}.
    \]
    Here we have used $N=\frac{n}{\log n}$.
    Using \autoref{prop:mult-ips-functional-lb}, any $\multips_{\Lin'}$ refutation of the unsatisfiable system $\set{f=0,X^2 - X=0,Y^2 - Y=0}$ computed by a formula needs size at-least $\Omega\inparen{\frac{n^2}{\log n}}$. 
    Note that both \autoref{lem:support-lem-1} and \autoref{obs:support-obs-1} are characteristic independent statement. 
    So, under the given partition $X_1\sqcup X_2\sqcup\cdots\sqcup X_N$, the set $\TL(\coeff_{X_i}(g))=\set{y_0,\ldots,y_{n-1}}$ remains unchanged for any $X_i$. 
    The algebraic independence of this set is independent of the characteristic of the field. 
    So, the lower bound works over characteristic $p$ as well.
\end{proof}
Next we show an example of a constant degree unsatisfiable system of equations such that any refutation in $\multips_{\Lin'}$ computed by formula needs near quadratic size. 
For that, first we recall polynomial $h'$ from \autoref{eq:poly1}.
Let $c>3$.
Further, let $X = \setdef{x_{i,j}}{i \in [c], j\in [n^2/c]}$ and $Y = \setdef{y_{i,j}}{i,j\in [n]}$ be sets of variables. 
The polynomial $h$ is set multilinear with respect to the partition $\bigsqcup_{i=1}^c\inparen{\bigcup_{k=1}^{n^{2(1-1/c)}/c} X_{i,k}}\bigsqcup Y$ where for any $i \in [c]$ and $k \in [n^{2(1-1/c)}/c]$, 
\[
    X_{i,k} = \setdef{x_{i,j}}{j \in \set{(k-1)(n^{2/c}) + 1 , \ldots, k (n^{2/c})}}
\]
\eqpolyone* 
Where each $h_k$ is set-multilinear over $\bigsqcup_{i=1}^c X_{i,k}\bigsqcup Y$ described in \autoref{eq:poly1}. Let $\bigsqcup_{i=1}^c X_{i,k}= X^{(k)}$ and $\inparen{\sqcup_{k=1}^{n^{2(1-1/c)}/c} X^{(k)}}\sqcup Y$ be a partition of variables. We use this partition to prove the $\multips_{\Lin'}$-lower bound against constant degree unsatisfiable systems. We first recall the polynomial.
\eqdefpolyone*
Next, we recall the theorem statement.
\formulaipslbconstantdeg*

\begin{proof}
    Let $g(X,Y)$ be the multilinear polynomial such that $g(X,Y)=\frac{1}{h(X,Y)}\mod{X^2 - X,Y^2 - Y}$.  
    For any two monomials $m_1,m_2\in \supp(h)$, $\supp(m_1)\nsubseteq\supp(m_2)\ \text{and}\ \supp(m_2)\nsubseteq\supp(m_1)$. 
    Hence \autoref{obs:support-obs-1} implies all monomials of $h$ appear in $g$ with a non-zero coefficient. Moreover, \autoref{lem:support-lem-1} implies every other monomial in $g$ with non-zero coefficient can only be the multilinearized product of monomials from $\supp(h)$. 
    
    Consider the partition of variables $\inparen{\bigsqcup_{k=1}^{n^{2(1-1/c)}/c} X^{(k)}}\bigsqcup Y$. 
    Consider the following monomial ordering: Over the variables, $X\succ Y$ and extend it to monomials naturally (\autoref{def:mon-order}). 
    Within variable set $X$ (and similarly $Y$), choose any order. 
    Under this ordering, $\TL(\coeff_{X^{(k)}}(g))=\setdef{Y_{i,j}}{i,j\in [n]}$ for every $X^{(k)}$ where $k\in [n^{2(1-1/c)}/c]$. All these variables are algebraically independent. 
    So, using \autoref{thm:alg-ind-LM,TM}, $\algrank_{X^{(k)}}(g)\geq n^2$. If $g$ has a formula of size $s$, then using \autoref{thm:kalorkoti}, $$s\geq \Omega\inparen{\sum_{i=1}^{n^{2(1-1/c)}/c}\algrank_{X^i}(g)+\algrank_{Y}(g)}\geq\Omega\inparen{ n^2\cdot n^{2(1-1/c)}/c+1}\geq\Omega\inparen{ n^{4-2/c}}.$$ 
    So, any formula computing $g$ needs size at least $\Omega(n^{4-2/c})$. 
    Hence, any $\multips_{\Lin'}$ refutation of $\set{h=0,X^2 - X=0,Y^2 - Y=0}$ computed by a formula needs size at-least $\Omega(n^{4-2/c})$ by \autoref{prop:mult-ips-functional-lb}. 
    Since both \autoref{lem:support-lem-1} and \autoref{obs:support-obs-1} is a characteristic independent statement, the set $\TL(\coeff_{X^{(k)}}(g))$ remains unchanged and algebraically independent. 
    So, the proof follows when characteristic of the field is positive.   
\end{proof} 

\section{Lower Bound Against Sum of ROABPs for $\ips_{\Lin'}$ Proofs}\label{sec:sum-roabp-ips-lin-lb}

We begin with an observation.
\begin{observation}\label{obs:mult-roabp}
   Let $f\in \mathbb{F}[x_1,\ldots,x_n]$ be a multilinear polynomial computed by an $\roabp$ of size $s$. Then there is a multilinear $\roabp$ of size at most $s$ computing the same polynomial.
\end{observation}
\begin{proof}
    Consider the $\roabp$, say $\abp$, computing $f$. 
    We get a multilinear $\roabp$ computing $f$ by simply removing non-multilinear terms from the label of every edge in $\abp$. 
    Since $f$ is multilinear, the contribution of the non-multilinear monomials in the edge labels anyway cancel out at the end and therefore this does not affect the computation of $f$.
\end{proof}

We next state a couple of theorems from the work of Forbes, Shpilka, Tzameret, and Wigderson \cite{FSTW21} that we will require.

\begin{lemma}\emph{(\cite[Lemma 3.7]{FSTW21})}\label{lem:ROABP_LB}
    Let $X = \set{x_1, \ldots, x_n}$ and $f \in \mathbb{F}[X]$ be a polynomial computed by an $\roabp$ of width $r$. 
    Then $r \geq \max_{i \in [n]} \set{\rank(M_{Y_i, Z_i}(f))}$ where $Y_i = \set{x_1, \ldots, x_i}$ and $Z_i = X \setminus Y_i$.     
\end{lemma}

\begin{lemma}\emph{(\cite[Proposition 4.5]{FSTW21})}\label{lem:ROABP-mult-ub} 
    Let $f\in \mathbb{F}[{X}]$ be a polynomial with individual degree of each variable being at most $d$.
    Further, suppose that $f$ is computable by an $\roabp$ of width $r$ in some order.
    Then $\mult(f)$ can be computed by an $\roabp$ of size $\poly(r,n,d)$ and width $r$ that has the same order.
    
    Further there exist polynomials $h_1, \cdots, h_n \in \mathbb{F}[X]$ such that 
    \begin{itemize}\itemsep0pt
        \item for every $i \in [n]$, $h_i$ has individual degree upper bounded by $d$;
        \item for every $i \in [n]$, $h_i$ can be computed by an $\roabp$ of size $\poly(r,n,d)$ and width $r$;
        \item $f(\vecx) = \mult(f) + \sum_{i=1}^n h_i(x_i^2-x_i)$. \qedhere
    \end{itemize}
 \end{lemma}
 
We now use \autoref{lem:ROABP-mult-ub} to show a similar statement for a sum of $\roabp$s as well.
 
\begin{proposition}[Multilinearization of Sum of ROABPs]\label{prop:sum-roabp-mult-ub} 
    Let $f\in \mathbb{F}[X]$ be a polynomial with individual degree at most $d$ such that it is computable by a sum of $t$ $\roabp$s, say $A_1, \ldots, A_t$, each with width at most $r$ and potentially a different variable ordering. 
    Let $\sigma_i$ be the variable ordering of $A_i$.
    Then,
    \begin{itemize}
        \item $\mult(f)$ is computable by a sum of $t$ multilinear $\roabp$s $B_1, \ldots, B_t$, where each $B_i$ has width at most $r$, size $\poly(r,n,d)$ and variable ordering $\sigma_i$.
        Further, if $f_i$ was the polynomial computed by $A_i$, then the polynomial computed by $B_i$ is $\mult(f_i)$;
        \item  there exist polynomials $h_1, \ldots, h_n$, each of individual degree at most $d$ and computable by a sum of $t$ $\roabp$s of size $\poly(r,n,d)$ and width at most $r$, such that 
        \[
            f(\vecx)=\mult(f)+\sum_{i=1}^n h_i(x_i^2-x_i).
        \]
    \end{itemize}
\end{proposition}

\begin{proof}
    We use \autoref{lem:ROABP-mult-ub} to prove the statement. 
    By the assumption of the lemma, $f$ is computable by $\sum_{i=1}^t A_i$ where each $A_i$ is an $\roabp$ of width at most $r$ and variable ordering $\sigma_i$.
    Let $f_i$ be the polynomial computed by $A_i$. 
    Note that the individual degree of $f_i$ is at most $d$. 
    
    Fix any $i$ arbitrarily. 
    Using \autoref{lem:ROABP-mult-ub}, $\mult(f_i)$ has an $\roabp$ of width at most $r$, order $\sigma_i$ and size $\poly(r,n,d)$. 
    Moreover, there exist polynomials $h_{i,1}, \ldots, h_{i,n}$ of individual degree at most $d$ such that $f_i = \mult(f_i) + \sum_{j=1}^n h_{i,j} (x_j^2 - x_j)$. 
    Here, for each $j \in [n]$, $h_{i,j}$ can be computed by an $\roabp$ of width at most $r$, order $\sigma_i$ and size $\poly(r,n,d)$. 
    Hence,
    \begin{align*}
        f = \sum_{i=1}^t f_i &= \sum_{i=1}^t \mult(f_i) + \sum_{i=1}^t \sum_{j=1}^n h_{i,j} (x_j^2-x_j)\\
        &= \sum_{i=1}^t \mult(f_i) + \sum_{j=1}^n \inparen{\sum_{i=1}^t h_{i,j}} (x_j^2 - x_j).
    \end{align*}
    Clearly $\sum_{i=1}^t \mult(f_i) = \mult(f)$. 
    Further, if we define $h_j = \sum_{i=1}^t h_{i,j}$, then each $h_j$ is computable by a sum of $t$ $\roabp$s that have the required properties.    
\end{proof}

We additionally require the following lemma from the work of Forbes, Shpilka, Tzameret, and Wigderson \cite{FSTW21}.

\begin{lemma}\emph{(\cite[Proposition 5.13]{FSTW21})}\label{lem:FSTW-HARD-poly}
    Let $\mathbb{F}$ be a field of characteristic zero, $X = \set{x_0, \ldots, x_{2n-1}}$, $T = \setdef{t_{i,j}}{i, j \in [0, \ldots, 2n-1] \text{ with } i< j}$ be two sets of variables and $\beta > \binom{2n}{2}$ be any number. 
    Further, let $g$ be the unique multilinear polynomial such that 
    \[
        g(X,T) \equiv \frac{1}{\sum_{i<j}t_{i,j}x_ix_j-\beta} \mod{X^2 - X, \ T^2 - T}.
    \]
    Then for any balanced partition $(Y, Z)$ of $X$, $\rank_{\mathbb{F}(T)}[M_{Y,Z}(g)] \geq 2^n$.
\end{lemma}

Finally, before we can prove our main theorem, we require a \emph{weakness lemma} for a sum of multilinear $\roabp$s. 
The proof of this lemma is based on ideas which are implicitly present in the work of Chatterjee, Kush, Saraf, and Shpilka \cite{CKSS24} (in the context of sums of ordered set-multilinear ABPs). 
However, we provide a detailed self-contained proof.

\subsection{Weakness of a Sum of Multilinear ROABPs}

Let $A = \sum_{i=1}^t A_i$ be a sum of $\roabp$s where each $A_i$ is multilinear and computing some multilinear polynomial in $\mathbb{F}[x_1,\ldots,x_n]$. 
Without loss of generality, we can assume that each $A_i$ has length $n$ and say the maximum width is $s$. 
Since the polynomial is multilinear, the partial derivative matrix under any partition of variables $(Y, Z)$ of $X$ has dimension $2^{\abs{Y}} \times 2^{\abs{Z}}$ \footnote{The rows and columns are indexed by multilinear monomials over $Y$ and $Z$ variables respectively.}.

\begin{fact}\label{fact:rank-up-pd-matrix} 
    Let $(Y,Z)$ be some partition of the variable set $X$ and $M_{Y,Z}(f)$ be the partial derivative matrix with respect to this partition.
    Then,
    \[
        \rank[M_{Y,Z}(f)] \leq \min\{2^{|Y|},2^{|Z|}\}=2^{\min\{|Y|,|Z|\}} \leq 2^{\frac{n-\abs{|Y|-|Z|}}{2}}.
    \]
\end{fact}
Next, we show that under any random balanced partition of the variables, the rank of any small size sum of multilinear $\roabp$ reduces significantly  with high probability.

\begin{restatable}[Weakness Lemma]{lemma}{sumROABPweakness}\label{lem:weakness-sum-roabp} 
    Let $q = r = \sqrt{n}$.
    Further, let $A = \sum_{i=1}^t A_i$ be a sum of multilinear $\roabp$s computing a polynomial in $\mathbb{F}[x_1,\ldots,x_n]$ and $(Y, Z)$ be a partition of the variable set $X$ chosen independently and uniformly at random. 
    Then, there exist constants $\epsilon',\epsilon''\in (0,1)$ such that 
    \[
        \Pr_{(Y,Z)} \insquare{\rank[M_{Y,Z}(A)] < t \cdot s^{q-1} \cdot 2^{\frac{n}{2}-\frac{\epsilon' q \sqrt{r}}{4}} \vert \ (Y,Z)\ \text{is balanced}} \geq 1-t\cdot e^{-\epsilon''q}.
    \]
\end{restatable}

\begin{proof}
    Since $(Y, Z)$ is a partition chosen independently and uniformly at random, each variable in $X$ is chosen to be a $Y$ variable with probability $1/2$ and a $Z$ variable with probability $1/2$. 
    Let this distribution on the partitions be $\mathcal{U}$. 
    A partition $(Y,Z)$ is balanced if $|Y|=|Z|$. 
    It is a standard fact \cite{CKSS24} that $\Pr_{(Y,Z) \sim \mathcal{U}} \insquare{(Y,Z) \text{ is balanced}} = \frac{\binom{n}{n/2}}{2^n} = \Theta \inparen{\frac{1}{\sqrt{n}}}$. 
    
    The idea is to first divide the $\roabp$s $\set{A_i}_{i \in [t]}$ into $q$ parts each of length $r$. 
    Fix $i \in [t]$ arbitrarily and let $u_0,u_q$ be the source and sink node of $A_i$. 
    Observe that
    \[
        A_i = \sum_{u_1,\ldots,u_{q-1}}\prod_{j=1}^q g_{u_{j-1}, u_j},
    \]
    where the node $u_j$ is in layer $j \in [q-1]$.
    Then the number of summands is upper bounded by $s^{q-1}$. 
    This division of the $\roabp$s naturally partitions the variable set $X = X_1 \sqcup \cdots \sqcup X_q$ with each $X_i$ containing $r$ variables. 
    $(Y,Z)$ also gets further partitioned naturally into $\set{(Y_j,Z_j)}_{j\in[q]}$. 
    We want to compute the rank of each product $\prod_{j=1}^q g_{u_{j-1},u_j}$ under any random partition.   
    \begin{multline*}
        \rank \insquare{M_{Y,Z} \inparen{\prod_{j=1}^{q} g_{u_{j-1},u_j}}} 
        = \prod_{j=1}^{q} \rank \insquare{M_{Y_j, Z_j}(g_{u_{j-1},u_j})} 
        \leq \prod_{j=1}^{q}2^{\frac{|Y_j|+|Z_j|}{2}-\frac{||Y_j|-|Z_j||}{2}}\\
        \leq 2^{\frac{n}{2}-\sum_{j=1}^{q}\frac{||Y_j|-|Z_j||}{2}}.
    \end{multline*}
    Hence if we can lower bound the term $\sum_{j=1}^{q}\frac{||Y_j|-|Z_j||}{2}$, we would be able to upper bound the rank. 
    Now, for any $j \in [q]$,
    \[
        \Pr_{(Y_j,Z_j)} \insquare{\frac{||Y_j|-|Z_j||}{2} \leq \frac{\sqrt{r}}{4}} 
        = \Pr_{(Y_j, Z_j)} \insquare{|Y_j|\in \insquare{\frac{r}{2}-\frac{\sqrt{r}}{4},\frac{r}{2}+\frac{\sqrt{r}}{4}}}
        =\sum_{k=\frac{r}{2}-\frac{\sqrt{r}}{4}}^{\frac{r}{2} +\frac{\sqrt{r}}{4}}\frac{\binom{r}{k}}{2^r}
        < 1.
    \]
    Let, 
    \[
        \epsilon = \sum_{k=\frac{r}{2}-\frac{\sqrt{r}}{4}}^{\frac{r}{2} +\frac{\sqrt{r}}{4}}\frac{\binom{r}{k}}{2^r} \text{\qquad and for any }  j \in [q], \text{ let \qquad}
        D_i = \begin{cases}
            1&\ \text{if }\ \frac{||Y_j|-|Z_j||}{2}\leq \frac{\sqrt{r}}{4}\\ 0&\ \text{otherwise}.
        \end{cases}
    \]
    Clearly, if $D=\sum_j D_j$, then $\mathbb{E}[D]=\sum_j\mathbb{E}[D_j] = \epsilon \cdot q$. 
    
    Fix $\delta \in (0,1)$ arbitrarily.
    Then, using Chernoff bound (\autoref{thm:chernoff-bound}), we know that 
    \[
        \Pr_{(Y,Z) \sim \mathcal{U}} [D \geq (1+\delta)\epsilon q]\leq \exp \inparen{-\frac{\delta^2\epsilon q}{2+\delta}}.
    \]
    Thus,
    \begin{multline*}
        \Pr_{(Y,Z)\sim \mathcal{U}}[D \geq (1+\delta)\epsilon q] \ | \ (Y,Z) \ \text{is balanced}] = \frac{\Pr_{(Y,Z) \sim \mathcal{U}} [D \geq (1+\delta)\epsilon q]}{\Pr_{(Y,Z)\sim \mathcal{U}} [(Y,Z) \ \text{is balanced}]}\\
                \leq \exp \inparen{-\frac{\delta^2\epsilon q}{2+\delta}} \cdot \sqrt{n}     
    \end{multline*}
    Choose $\epsilon'' = \inparen{\frac{2\delta^2\epsilon}{2+\delta}}$ and $\hat{\epsilon} = \epsilon(1+\delta)$. Then, we have that 
    \begin{multline*}
            \Pr_{(Y,Z)\sim \mathcal{U}}[D \geq \hat{\epsilon} q] \ | \ (Y,Z)\ \text{is balanced}] \leq \exp \inparen{-\epsilon'' q}\\
            \implies \Pr_{(Y,Z)\sim \mathcal{U}}[D < \hat{\epsilon} q\ | \ (Y,Z)\ \text{is balanced}] \geq 1- \exp \inparen{-\epsilon'' q}\\
            \implies \Pr_{(Y,Z)\sim \mathcal{U}} \insquare{\sum_{j=1}^q \frac{||Y_j|-|Z_j||}{2} > \frac{(1-\hat{\epsilon})q\sqrt{r}}{4}\ |\ (Y,Z)\ \text{is balanced}}\geq 1-\exp(-\epsilon''q)\\
            \implies \Pr_{(Y,Z)\sim \mathcal{U}} \insquare{\rank \inparen{M_{Y,Z}\inparen{\prod_{j=1}^{q} g_{u_{j-1},u_j}}} < 2^{\frac{n}{2}-\frac{(1-\hat{\epsilon})q\sqrt{r}}{4}}\ \vert \ (Y,Z)\ \text{is balanced}}\\
            \geq 1-\exp(-\epsilon''q). 
    \end{multline*}
    Recall that $A_i = \sum_{u_1,\ldots,u_{q-1}}\prod_{j=1}^q g_{u_{j-1}, u_j}$. 
    Thus, for every $i \in [t]$,
    \[
        \Pr_{(Y,Z)\sim \mathcal{U}} \insquare{\rank \inparen{M_{Y,Z}\inparen{A_i}} < s^{q-1} \cdot 2^{\frac{n}{2}-\frac{(1-\hat{\epsilon})q\sqrt{r}}{4}}\ \vert \ (Y,Z)\ \text{is balanced}} \geq 1-\exp(-\epsilon''q).
    \]
    By union bound, this shows that 
    \begin{equation*}
        \begin{split}
            \Pr_{(Y,Z)\sim \mathcal{U}} \insquare{\exists i \in [t] \text{ s.t. } \rank \inparen{M_{Y,Z}\inparen{A_i}} \geq s^{q-1} \cdot 2^{\frac{n}{2} - \frac{(1-\hat{\epsilon})q\sqrt{r}}{4}}\ \vert \ (Y,Z)\ \text{is balanced}}\\
            \leq t \cdot \exp(-\epsilon''q)\\
            \implies \Pr_{(Y,Z)\sim \mathcal{U}} \insquare{\forall i \in [t], \ \rank \inparen{M_{Y,Z}\inparen{A_i}} < s^{q-1} \cdot 2^{\frac{n}{2} - \frac{(1-\hat{\epsilon})q\sqrt{r}}{4}}\ \vert \ (Y,Z)\ \text{is balanced}}\\
            \geq 1 - t \cdot \exp(-\epsilon''q).
        \end{split}
    \end{equation*}
    Finally, using the sub-additivity of rank, we get that
    \[
        \Pr_{(Y,Z)\sim \mathcal{U}} \insquare{\rank \inparen{M_{Y,Z}\inparen{A}} < t \cdot s^{q-1} \cdot 2^{\frac{n}{2} - \frac{(1-\hat{\epsilon})q\sqrt{r}}{4}}\ \vert \ (Y,Z)\ \text{is balanced}} \geq 1 - t \cdot \exp(-\epsilon''q).
    \]
    Choosing $\epsilon' = 1 - \hat{\epsilon}$ completes the proof.
\end{proof}

We are now ready to prove an exponential lower bound against $\sum \roabp \text{-} \ips_{\Lin'}$ proofs.

\subsection{Lower Bound over Fields of Characteristic Zero}
\sumofroabp*
\begin{proof}
    Let $g$ be the unique multilinear polynomial that proves the unsatisfiability of $f$. 
    Then, by \autoref{lem:FSTW-HARD-poly}, for any balanced partition $(Y, Z)$ of $X$, $\rank_{\mathbb{F}(T)}[M_{Y,Z}(g)]\geq 2^n$. 
    Thus,
    \[
        \Pr_{(Y,Z)} \insquare{\rank_{\mathbb{F}(T)}[M_{Y,Z}(g)] < 2^n \vert \ (Y,Z)\ \text{is balanced}} = 0.
    \]
    
    Suppose $g$ is computable by a sum of $t$ multilinear $\roabp$s, say $A_1, \ldots A_t$, of width at most $s$. 
    Since the total number of variables is $2n$, by \autoref{lem:weakness-sum-roabp}, we have that for any partition $(Y,Z)$ of $X$ chosen uniformly at random,
    \[
        \Pr_{(Y,Z)} \insquare{\rank_{\mathbb{F}(T)}[M_{Y,Z}(g)] < t \cdot s^{q-1} \cdot 2^{n-\frac{\epsilon' q \sqrt{r}}{4}} \vert \ (Y,Z)\ \text{is balanced}} \geq 1-t\cdot e^{-\epsilon''q}.
    \]
    for suitable constants $\epsilon', \epsilon'' \in (0,1)$ and $q = r = \sqrt{2n}$. 
    
    Note that if $t> \exp(\epsilon''q)$, then we already have an $\exp(\Omega(\sqrt{n}))$ lower bound.
    Otherwise, there exists a balanced partition $(Y,Z)$ such that $\rank_{\mathbb{F}(T)}[M_{Y,Z}(g)] \leq t \cdot s^{q-1}\cdot 2^{n-\frac{\epsilon' q \sqrt{r}}{4}}$. 
    However, we know that $\rank_{\mathbb{F}(T)}[M_{Y,Z}(g)] \geq 2^n$. 
    Hence, it must be the case that $t \cdot s^{q-1} \geq 2^{\epsilon' q \sqrt{r}}$, which would imply that $s \geq \exp(n^{1/4})$. 
    Either way, we get that the size of the sum of $\roabp$s computing $g$ is at least $\exp(n^{\gamma})$ for some $\gamma >0$.
    
    Let $h \in \ips_{\Lin,}$ be any proof of unsatisfiability for $f$.
    Then $\mult(h)=g$. 
    Assume, for sake of contradiction, that $h$ can be computed by a sum of $\roabp$s, say $\sum A_i$, of size $\exp(o(n^\gamma))$. 
    \[
        h = \sum_i A_i \implies \mult(h) = \mult \inparen{\sum_i A_i} = \sum_i \mult(A_i)
    \]
    Each $\roabp$ $\mult(A_i)$ is a multilinear $\roabp$ and using \autoref{prop:sum-roabp-mult-ub}, $\sum_i \mult(A_i)$ is of size $\exp(o(n^{\gamma'}))$, but this contradicts the fact that any sum of multilinear $\roabp$ computing $\mult(h)=g$ needs size at least $\exp(\Omega(n^\gamma))$. 
    This completes the proof.
\end{proof}

\subsection{Lower Bounds over Fields of Positive Characteristic of Large Size}\label{sec:sum-of roabp-ips-lb-positive-char}

Note that in \autoref{thm:sum-roabp-ips-lb}, the characteristic of $\mathbb{F}$ is required to be zero even though the weakness lemma (\autoref{lem:weakness-sum-roabp}) for sum of multilinear-$\roabp$s is characteristic independent.
This is because we use \autoref{lem:FSTW-HARD-poly}. 
So, in order to prove a theorem analogous to \autoref{thm:sum-roabp-ips-lb} in the positive characteristic setting, we need a statement analogous to \autoref{lem:FSTW-HARD-poly} in this setting.

The main idea for such a rank lower bound 
was recently shown in a work of Behera, Limaye, Ramanathan and Srinivasan \cite{BLRS25}. 
We first state the main lemma from their work.

\begin{lemma}\emph{(\cite[Lemma 2.4]{BLRS25})}\label{lem:deg-bound-positive-char} 
    Let $\mathbb{F},\mathbb{F}'$ be two fields such that $\mathbb{F}\subset \mathbb{F}'$, $n\in \mathbb{N}$ and $X$ be the variable set. 
    Fix $\beta \in \mathbb{F}'\backslash \mathbb{F}$ arbitrarily. 
    Further, for any $\bar{\alpha}\in \mathbb{F}^n$ and a non-empty subset $S' \subseteq [n]$, let $g_{\bar{\alpha},S'}(\vecx) \in \mathbb{F}[X]$ be the unique multilinear polynomial that agrees with $\frac{1}{\sum_{i\in S'}\alpha_i\cdot x_i-\beta}$ over Boolean hypercube.
    
    For any $S\subseteq \mathbb{F}$ which is finite, if we choose $\bar{\alpha}$ uniformly at random from $S^n$, then the following is true.
    \[
        \Pr_{\bar{\alpha}\sim S^n} \insquare{\exists \ \emptyset\neq S'\subseteq [n]\ :\ \deg(g_{\bar{\alpha},S'}(\vecx))<|S'|} < \frac{2^{2n}}{|S|} \qedhere
    \]
\end{lemma}

\noindent We will also need the following lemma from the work of Forbes, Shpilka, Tzameret, and Wigderson \cite{FSTW21}.

\begin{lemma}\emph{(\cite[Lemma 5.12]{FSTW21})}\label{lem:rank-coef-lemma}
    Let $f\in \mathbb{F}[X,Y,Z]$ and $f_Z \in \mathbb{F}(Z)[X,Y]$ be the polynomial that symbolically equals $f$. 
    That is, for any $\gamma\in \mathbb{F}^{|Z|}$, we have $f_{\gamma}(X,Y)= f(X,Y,\gamma) \in \mathbb{F}[X,Y]$. 
    
    For any set of variables $X$, any field $\mathbb{F}$ and any $f \in \mathbb{F}[X]$, suppose $\coeff_{X}(f)$ is used to denote the coefficient vector of $f$.
    Then, for any $\gamma \in \mathbb{F}^{|Z|}$, $\dim_{\mathbb{F}(Z)}[\coeff_{X|Y}[f_Z(X,Y)]]\geq \dim_{\mathbb{F}}[\coeff_{X|Y}[f_{\gamma}(X,Y)]]$.
\end{lemma}

Using \autoref{lem:deg-bound-positive-char} and \autoref{lem:rank-coef-lemma}, Behera, Limaye, Ramanathan and Srinivasan \cite{BLRS25} prove a rank bound analogous to \autoref{lem:FSTW-HARD-poly} in positive characteristic.

\begin{restatable}{lemma}{rankboundpositivechar}\emph{(\cite[Lemma A.10]{BLRS25})}\label{thm:rank-bound-positive-char}  
    Let $n\in \mathbb{N}$ and $p\in \mathbb{N}$ be any prime. 
    Say $\mathbb{F}'$ is a field of characteristic $p$ with size $p^{k+1}$, where $k$ is the smallest integer such that $p^k> \binom{2n}{n}2^{2n}$ and that $\mathbb{F}$ is a field of size $p^k$ (so that $\mathbb{F}\subset \mathbb{F}'$).
    Fix $\beta\in \mathbb{F}'\backslash \mathbb{F}$ arbitrarily and finally, for any ${\alpha}\in \mathbb{F}^{\binom{2n}{2}}$, let $g_{\alpha}(\vecx,\vect) \in \mathbb{F}'[X,T]$ be the  polynomial that agrees with $\frac{1}{\sum_{i<j}\alpha_{i,j}t_{i,j}x_ix_j-\beta}$ on the Boolean hypercube. 
    
    Then there exists ${\bar{\alpha}}$ such that for any balanced partition $(U,V)$ of $X$,
    $\rank_{\mathbb{F}'(T)}[M_{U,V}(g_{{\alpha}})] \geq 2^n.$  
\end{restatable}

We are now ready to prove an exponential lower bound for sum of $\roabp$s in positive characteristic.

\begin{restatable}{theorem}{sumofroabppositivechar}\label{thm:sum-roabp-ips-lb-postive-char}
    Let $n\in \mathbb{N}$ and $p$ be any prime. 
    Further let $\mathbb{F}'$ be a field of characteristic $p$ with size $p^{k+1}$, where $k$ is the smallest integer such that $p^k> \binom{2n}{2}2^{2n}$ and $\mathbb{F} \subset \mathbb{F}'$ be the subfield of size $p^k$.
    Also, arbitrarily fix $\beta\in \mathbb{F}'\backslash \mathbb{F}$. 
    
    For $X = \set{x_0, \ldots, x_{2n-1}}$, $T = \setdef{t_{i,j}}{i, j \in \set{0, \ldots, 2n-1 \text{ with } i<j}}$ and any $\bar{\alpha}\in \mathbb{F}^{\binom{2n}{2}}$, define
    \[
        f = \inparen{\sum_{0 \leq i < j \leq 2n-1}\alpha_{i,j} t_{i,j} x_ix_j} - \beta
    \]
    which is unsatisfiable over the Boolean hypercube.
    Then there exists $\bar{\alpha}$ and $\gamma>0,$ such that any sum of $\roabp$ computing the $\ips_{\Lin'}$ refutation of the system $\{f = 0, X^2 - X = 0,T^2-T=0\}$ must have total width at-least $\exp(n^\gamma)$.
\end{restatable}

\begin{proof}
 We combine \autoref{thm:rank-bound-positive-char} with the weakness lemma (\autoref{lem:weakness-sum-roabp}) to get the required lower bound. The argument works in a similar manner given in the proof of \autoref{thm:sum-roabp-ips-lb}.
\end{proof}

\subsection{Lower Bound over Fields of Characteristic At Least Five}

In \autoref{thm:sum-roabp-ips-lb-postive-char} the field size needs to be large since we are using \autoref{lem:deg-bound-positive-char} to prove the rank lower bound. 
However, we can remove such a size requirement if we use the vector invariant polynomial from the work of Hakoniemi, Limaye, and Tzameret \cite{HLT24}. 
We first state the rank lower bound from their work.

\begin{lemma}\emph{(\cite[Lemma 43]{HLT24})}\label{lem:rank-lb-finite-field}
    Let $\mathbb{F}$ be a field of characteristic at least and $g(\vecx, \vect) \in \mathbb{F}[X,T]$ be the polynomial defined in \autoref{thm:sum-roabp-ips-lb-positive-char-small-field}. 
    Let $\hat{g}(\vecx, \vect) \in \mathbb{F}[X,T]$ be any polynomial that satisfies 
    \[
        \hat{g}(\vecx, \vect) = \frac{1}{g(\vecx, \vect)} \bmod(X^2 - X,T^2-T).
    \]
    If $\hat{g}_T(\vecx) \in \mathbb{F}(T)[X]$ is the polynomial that symbolically equals $\hat{g}$, then for any balanced partition $(U,V)$ of $X$, $\rank_{\mathbb{F}(T)}[M_{U,V}(\hat{g}_T(X)]\geq 2^n$.
\end{lemma}

We are now ready to prove an exponential lower bound against $\sum$ $\roabp$s for proofs in $\ips_{\Lin'}$ model when the field size is not necessarily large. 

\begin{restatable}{theorem}{sumofroabpfinitefield}\label{thm:sum-roabp-ips-lb-positive-char-small-field}
    Let $\mathbb{F}$ be a field of characteristic at least $5$.
    Further, let $X=\{x_1,\ldots,x_{4n}\}$ and $T = \setdef{t_{i,j,k,l}}{i,j,k,l \in [4n] \text{ with } i<j<k<l}$ be two sets of variables. 
    
    Define $g\in \mathbb{F}[X,T]$ to be the polynomial 
    \[
        g=\inparen{\prod_{1 \leq i<j<k<l \leq 4n} 1-t_{i,j,k,l}+t_{i,j,k,l}(x_ix_l-x_jx_k)}-\beta
    \]
    which is unsatisfiable over Boolean hypercube as long as $\beta \neq \{-1,0,1\}$.  
   Then there exists $\gamma>0$ such that any sum of $\roabp$ computing a $\ips_{\Lin'}$ refutation of the system $\{g = 0, X^2 - X = 0,T^2-T=0\}$ must have total width at-least $\exp(n^\gamma)$.
\end{restatable}

The proof follows exactly along the lines of the one for \autoref{thm:sum-roabp-ips-lb-postive-char} except that we use \autoref{lem:weakness-sum-roabp} combined with \autoref{lem:rank-lb-finite-field} instead of \autoref{thm:rank-bound-positive-char}.

\section*{Conclusion}\label{sec:conclusion}

The construction of the axiom polynomials from the subset-sum axiom polynomial is particularly relevant in proof complexity although the subset-sum polynomial is not directly a translation of CNFs. More details regarding this can be found in \cite[Section 1.3]{GHT22}.
If the axiom polynomial is sparse, then it can be directly constructed from the subset-sum polynomial by substituting monomials for the variables. In this paper, all the axiom polynomials except the one used in \autoref{thm:sum-roabp-ips-lb-positive-char-small-field} are sparse. 

However, if one does not care about constructing sparse axiom polynomials, then the problem of proving polynomial-size quality lower bounds for formulas, ABPs, and circuits can be solved easily. This is noted in \autoref{formula-ips-mult-obs}. As already mentioned above, it is more desirable if $\ips$ lower bound results are shown for sparse axioms. This is also reflected if one compares the work of \cite{AF22} with \cite{GHT22}. 

Finally, we state a few open problems for further study. 

\begin{enumerate}
\item Prove (nearly) quadratic-size $\ips_{\Lin'}$ lower bounds for formulas for an axiom polynomial which is sparse. 

\item Prove super-linear size $\multips_{\Lin'}$ and $\ips_{\Lin'}$ lower bounds for ABPs and circuits. Again, the axiom polynomials should be sparse.  

\item Can we (re)-prove \autoref{thm:sum-roabp-ips-lb-positive-char-small-field} for a sparse axiom polynomial? 
\end{enumerate}

\bibliographystyle{customurlbst/alphaurlpp}
\bibliography{ref}

\newcommand{\etalchar}[1]{$^{#1}$}
\begin{thebibliography}{FSTW21}

\bibitem[AF22]{AF22}
Robert Andrews and Michael~A. Forbes.
\newblock \href {http://dx.doi.org/10.1145/3519935.3520025} {Ideals,
  determinants, and straightening: proving and using lower bounds for
  polynomial ideals}.
\newblock In {\em {STOC} '22: 54th Annual {ACM} {SIGACT} Symposium on Theory of
  Computing, Rome, Italy, June 20 - 24, 2022}, pages 389--402. {ACM}, 2022.

\bibitem[Ajt88]{A88}
Mikl{\'o}s Ajtai.
\newblock \href {http://dx.doi.org/10.1109/SFCS.1988.21948} {The complexity of
  the pigeonhole principle}.
\newblock In {\em 29th Annual Symposium on Foundations of Computer Science
  (FOCS)}, pages 346--355. IEEE, 1988.

\bibitem[AR01]{AR01}
Michael Alekhnovich and Alexander~A. Razborov.
\newblock \href {http://dx.doi.org/10.1109/SFCS.2001.959893} {Lower Bounds for
  Polynomial Calculus: Non-Binomial Case}.
\newblock In {\em 42nd Annual Symposium on Foundations of Computer Science,
  {FOCS} 2001, Las Vegas, Nevada, USA, October 14-17, 2001}, pages 190--199.
  {IEEE} Computer Society, 2001.

\bibitem[BCH94]{BCH1994}
Paul Beame, Stephen~A. Cook, and H.~James Hoover.
\newblock \href {http://dx.doi.org/10.1137/S0097539791197240} {Log Depth
  Circuits for Division and Related Problems}.
\newblock {\em SIAM Journal on Computing}, 23(4):740--751, 1994.

\bibitem[BCS13]{BCS13}
Peter B{\"u}rgisser, Michael Clausen, and Mohammad~A Shokrollahi.
\newblock {\em Algebraic complexity theory}, volume 315.
\newblock Springer Science \& Business Media, 2013.

\bibitem[BDS25]{BDS25}
C.~S. Bhargav, Prateek Dwivedi, and Nitin Saxena.
\newblock \href {http://dx.doi.org/10.1016/J.TCS.2025.115214} {Lower bounds for
  the sum of small-size algebraic branching programs}.
\newblock {\em Theor. Comput. Sci.}, 1041:115214, 2025.

\bibitem[BGIP01]{BGIP01}
Samuel~R. Buss, Dima Grigoriev, Russell Impagliazzo, and Toniann Pitassi.
\newblock \href {http://dx.doi.org/10.1006/JCSS.2000.1726} {Linear Gaps between
  Degrees for the Polynomial Calculus Modulo Distinct Primes}.
\newblock {\em J. Comput. Syst. Sci.}, 62(2):267--289, 2001.

\bibitem[BIK{\etalchar{+}}94]{BIKPP94}
Paul Beame, Russell Impagliazzo, Jan Kraj{\'{\i}}cek, Toniann Pitassi, and
  Pavel Pudl{\'{a}}k.
\newblock \href {http://dx.doi.org/10.1109/SFCS.1994.365714} {Lower Bound on
  Hilbert's Nullstellensatz and propositional proofs}.
\newblock In {\em 35th Annual Symposium on Foundations of Computer Science,
  Santa Fe, New Mexico, USA, 20-22 November 1994}, pages 794--806. {IEEE}
  Computer Society, 1994.

\bibitem[BIK{\etalchar{+}}97]{BPRS97}
Samuel~R. Buss, Russell Impagliazzo, Jan Kraj{\'{\i}}cek, Pavel Pudl{\'{a}}k,
  Alexander~A. Razborov, and Jir{\'{\i}} Sgall.
\newblock \href {http://dx.doi.org/10.1007/BF01294258} {Proof Complexity in
  Algebraic Systems and Bounded Depth Frege Systems with Modular Counting}.
\newblock {\em Comput. Complex.}, 6(3):256--298, 1997.

\bibitem[BLRS25]{BLRS25}
Amik~Raj Behera, Nutan Limaye, Varun Ramanathan, and Srikanth Srinivasan.
\newblock \href {https://eccc.weizmann.ac.il/report/2025/079/} {New Bounds for
  the Ideal proof System in Positive Characteristic}.
\newblock {\em Electron. Colloquium Comput. Complex.}, TR25-079, 2025.
\newblock Pre-print available at \href {http://arxiv.org/abs/TR25-O79}
  {\path{arXiv:TR25-O79}}.

\bibitem[BO83]{b83}
Michael Ben-Or.
\newblock \href {http://dx.doi.org/10.1145/800061.808735} {Lower Bounds for
  Algebraic Computation Trees}.
\newblock In {\em Proceedings of the Fifteenth Annual ACM Symposium on Theory
  of Computing (STOC)}, pages 80--86. ACM, 1983.

\bibitem[BPI92]{BPI93}
Paul Beame, Toniann Pitassi, and Russell Impagliazzo.
\newblock \href {http://dx.doi.org/10.1145/129712.129739} {Exponential lower
  bounds for the pigeonhole principle}.
\newblock In {\em 24th Annual ACM Symposium on Theory of Computing (STOC)},
  pages 200--220. ACM, 1992.

\bibitem[BS83]{BS83}
Walter Baur and Volker Strassen.
\newblock The Complexity of Partial Derivatives.
\newblock {\em Theor. Comput. Sci.}, 22:317--330, 1983.

\bibitem[CKSS24]{CKSS24}
Prerona Chatterjee, Deepanshu Kush, Shubhangi Saraf, and Amir Shpilka.
\newblock \href {http://dx.doi.org/10.4230/LIPICS.CCC.2024.20} {Lower Bounds
  for Set-Multilinear Branching Programs}.
\newblock In {\em 39th Computational Complexity Conference, {CCC} 2024, July
  22-25, 2024, Ann Arbor, MI, {USA}}, volume 300 of {\em LIPIcs}, pages
  20:1--20:20. Schloss Dagstuhl - Leibniz-Zentrum f{\"{u}}r Informatik, 2024.

\bibitem[CKSV22]{CKSV22}
Prerona Chatterjee, Mrinal Kumar, Adrian She, and Ben~Lee Volk.
\newblock \href {http://dx.doi.org/10.1007/S00037-022-00223-8} {Quadratic Lower
  Bounds for Algebraic Branching Programs and Formulas}.
\newblock {\em Comput. Complex.}, 31(2):8, 2022.

\bibitem[CKV24]{CKV24}
Abhranil Chatterjee, Mrinal Kumar, and Ben~Lee Volk.
\newblock \href {http://dx.doi.org/10.1007/S00037-024-00258-Z} {Determinants
  vs. Algebraic Branching Programs}.
\newblock {\em Comput. Complex.}, 33(2):11, 2024.

\bibitem[CR79]{CR79}
Stephen~A. Cook and Robert~A. Reckhow.
\newblock \href {http://dx.doi.org/10.2307/2273702} {The Relative Efficiency of
  Propositional Proof Systems}.
\newblock {\em Journal of Symbolic Logic}, 44(1):36--50, 1979.

\bibitem[EGLT25]{EGLT25}
Tal Elbaz, Nashlen Govindasamy, Jiaqi Lu, and Iddo Tzameret.
\newblock \href {https://arxiv.org/abs/2506.17210} {Lower Bounds against the
  Ideal Proof System in Finite Fields}.
\newblock {\em arXive Preprint}, June 2025.
\newblock Preprint.
\newblock Pre-print available at \href {http://arxiv.org/abs/2506.17210}
  {\path{arXiv:2506.17210}}.

\bibitem[For24]{F24}
Michael~A. Forbes.
\newblock \href {http://dx.doi.org/10.4230/LIPICS.CCC.2024.31} {Low-Depth
  Algebraic Circuit Lower Bounds over Any Field}.
\newblock In {\em 39th Computational Complexity Conference, {CCC} 2024, July
  22-25, 2024, Ann Arbor, MI, {USA}}, volume 300 of {\em LIPIcs}, pages
  31:1--31:16. Schloss Dagstuhl - Leibniz-Zentrum f{\"{u}}r Informatik, 2024.

\bibitem[FS15]{FS15}
Michael~A. Forbes and Amir Shpilka.
\newblock \href {http://dx.doi.org/10.1145/2852040.2852051} {Complexity Theory
  Column 88: Challenges in Polynomial Factorization1}.
\newblock {\em {SIGACT} News}, 46(4):32--49, 2015.

\bibitem[FSTW21]{FSTW21}
Michael~A. Forbes, Amir Shpilka, Iddo Tzameret, and Avi Wigderson.
\newblock \href {https://theoryofcomputing.org/articles/v017a010/} {Proof
  Complexity Lower Bounds from Algebraic Circuit Complexity}.
\newblock {\em Theory Comput.}, 17:1--88, 2021.

\bibitem[GHT22]{GHT22}
Nashlen Govindasamy, Tuomas Hakoniemi, and Iddo Tzameret.
\newblock \href {http://dx.doi.org/10.1109/FOCS54457.2022.00025} {Simple Hard
  Instances for Low-Depth Algebraic Proofs}.
\newblock In {\em 63rd {IEEE} Annual Symposium on Foundations of Computer
  Science, {FOCS} 2022, Denver, CO, USA, October 31 - November 3, 2022}, pages
  188--199. {IEEE}, 2022.

\bibitem[GP18]{GP18}
Joshua~A. Grochow and Toniann Pitassi.
\newblock \href {http://dx.doi.org/10.1145/3230742} {Circuit Complexity, Proof
  Complexity, and Polynomial Identity Testing: The Ideal Proof System}.
\newblock {\em J. {ACM}}, 65(6):37:1--37:59, 2018.

\bibitem[H{\aa}s23]{H23}
Johan H{\aa}stad.
\newblock \href {http://dx.doi.org/10.1109/FOCS57990.2023.00010} {On
  small-depth Frege proofs for {PHP}}.
\newblock In {\em 64th {IEEE} Annual Symposium on Foundations of Computer
  Science, {FOCS} 2023, Santa Cruz, CA, USA, November 6-9, 2023}, pages 37--49.
  {IEEE}, 2023.

\bibitem[HLT24]{HLT24}
Tuomas Hakoniemi, Nutan Limaye, and Iddo Tzameret.
\newblock \href {http://dx.doi.org/10.1145/3618260.3649616} {Functional Lower
  Bounds in Algebraic Proofs: Symmetry, Lifting, and Barriers}.
\newblock In {\em Proceedings of the 56th Annual {ACM} Symposium on Theory of
  Computing, {STOC} 2024, Vancouver, BC, Canada, June 24-28, 2024}, pages
  1396--1404. {ACM}, 2024.

\bibitem[HR22]{HR22}
Johan H{\aa}stad and Kilian Risse.
\newblock \href {http://dx.doi.org/10.1109/FOCS54457.2022.00110} {On Bounded
  Depth Proofs for Tseitin Formulas on the Grid; Revisited}.
\newblock In {\em 63rd {IEEE} Annual Symposium on Foundations of Computer
  Science, {FOCS} 2022, Denver, CO, USA, October 31 - November 3, 2022}, pages
  1138--1149. {IEEE}, 2022.

\bibitem[HY09]{HY09}
Pavel Hrubes and Amir Yehudayoff.
\newblock \href {http://dx.doi.org/10.1016/J.IPL.2009.09.003} {Monotone
  separations for constant degree polynomials}.
\newblock {\em Inf. Process. Lett.}, 110(1):1--3, 2009.

\bibitem[IPS99]{IPS99}
Russell Impagliazzo, Pavel Pudl{\'{a}}k, and Jir{\'{\i}} Sgall.
\newblock \href {http://dx.doi.org/10.1007/S000370050024} {Lower Bounds for the
  Polynomial Calculus and the Gr{\"{o}}bner Basis Algorithm}.
\newblock {\em Comput. Complex.}, 8(2):127--144, 1999.

\bibitem[Kal85]{kal}
K.~A. Kalorkoti.
\newblock \href {http://dx.doi.org/10.1137/0214050} {A Lower Bound for the
  Formula Size of Rational Functions}.
\newblock {\em SIAM Journal on Computing}, 14(3):678--687, 1985.

\bibitem[KPW95]{KPW95}
Jan Krajíček, Pavel Pudlák, and Alan Woods.
\newblock \href {http://dx.doi.org/10.1002/rsa.3240070103} {Exponential lower
  bounds to the size of bounded depth Frege proofs of the pigeonhole
  principle}.
\newblock {\em Random Structures \& Algorithms}, 7(1):15--39, 1995.

\bibitem[KR05]{KR05}
Martin Kreuzer and Lorenzo Robbiano.
\newblock \href {http://dx.doi.org/10.1007/b138062} {{\em Computational
  Commutative Algebra 2}}, volume~2 of {\em Springer-Verlag Algorithms and
  Computation in Mathematics}.
\newblock Springer-Verlag, Berlin, Heidelberg, 2005.

\bibitem[LST21]{LST21}
Nutan Limaye, Srikanth Srinivasan, and S{\'{e}}bastien Tavenas.
\newblock \href {https://doi.org/10.1109/FOCS52979.2021.00083} {Superpolynomial
  Lower Bounds Against Low-Depth Algebraic Circuits}.
\newblock In {\em 62nd {IEEE} Annual Symposium on Foundations of Computer
  Science, {FOCS} 2021, Denver, CO, USA, February 7-10, 2022}, pages 804--814.
  {IEEE}, 2021.

\bibitem[LTW18]{LTW18}
Fu~Li, Iddo Tzameret, and Zhengyu Wang.
\newblock \href {http://dx.doi.org/10.1137/16M1107632} {Characterizing
  Propositional Proofs as Noncommutative Formulas}.
\newblock {\em {SIAM} J. Comput.}, 47(4):1424--1462, 2018.

\bibitem[PI94]{PI1994}
Toniann Pitassi and Russell Impagliazzo.
\newblock \href {http://dx.doi.org/10.1145/195058.195120} {Exponential Lower
  Bounds for the Polynomial Calculus}.
\newblock In {\em Proceedings of the 26th Annual ACM Symposium on Theory of
  Computing (STOC)}, pages 129--135. ACM, 1994.

\bibitem[Pit96]{Pit96}
Toniann Pitassi.
\newblock \href {http://dx.doi.org/10.1090/DIMACS/031/07} {Algebraic
  Propositional Proof Systems}.
\newblock In {\em Descriptive Complexity and Finite Models, Proceedings of a
  {DIMACS} Workshop 1996, Princeton, New Jersey, USA, January 14-17, 1996},
  volume~31 of {\em {DIMACS} Series in Discrete Mathematics and Theoretical
  Computer Science}, pages 215--244. {DIMACS/AMS}, 1996.

\bibitem[PRST16]{PRST16}
Toniann Pitassi, Benjamin Rossman, Rocco~A. Servedio, and Li{-}Yang Tan.
\newblock \href {http://dx.doi.org/10.1145/2897518.2897637} {Poly-logarithmic
  Frege depth lower bounds via an expander switching lemma}.
\newblock In {\em Proceedings of the 48th Annual {ACM} {SIGACT} Symposium on
  Theory of Computing, {STOC} 2016, Cambridge, MA, USA, June 18-21, 2016},
  pages 644--657. {ACM}, 2016.

\bibitem[Raz98]{R98}
Alexander~A. Razborov.
\newblock \href {http://dx.doi.org/10.1007/S000370050013} {Lower Bounds for the
  Polynomial Calculus}.
\newblock {\em Comput. Complex.}, 7(4):291--324, 1998.

\bibitem[Rec76]{R76}
Robert~A. Reckhow.
\newblock {\em On the lengths of proofs in the propositional calculus}.
\newblock Ph.d.\ thesis, University of Toronto, 1976.
\newblock Technical Report \#87.

\bibitem[Sap21]{rpsurvey}
Ramprasad Saptharishi.
\newblock A survey of lower bounds in arithmetic circuit complexity.
\newblock Manuscript online at
  https://github.com/dasarpmar/lowerbounds-survey/releases/download/v9.0.3/fancymain.pdf,
  2021.
\newblock A selection of lower bounds in arithmatic circuit complexity.

\bibitem[SW01]{SW01}
Amir Shpilka and Avi Wigderson.
\newblock \href {http://dx.doi.org/10.1007/PL00001609} {Depth--3 Arithmetic
  Circuits Over Fields of Characteristic Zero}.
\newblock {\em Computational Complexity}, 10(1):1--27, November 2001.

\bibitem[SY10]{SY10}
Amir Shpilka and Amir Yehudayoff.
\newblock \href {http://dx.doi.org/10.1561/0400000039} {Arithmetic Circuits:
  {A} survey of recent results and open questions}.
\newblock {\em Found. Trends Theor. Comput. Sci.}, 5(3-4):207--388, 2010.

\end{thebibliography}

\appendix \label{appendix}

\section{Some Related Observations}\label{sec:IPS-UB}
In this section, we first sketch the non-multilinear formula refutations upper bounds. 
We start by defining the subset-sum polynomial.
\begin{definition}
    Let $X=\set{x_1,\ldots,x_n}$ be a set of variables and $\beta$ be a constant from $\mathbb{F}$. Then the subset-sum polynomial $f\in \mathbb{F}[X]$ is the following,
    \begin{restatable}{restatableeqn}{eqnpolythree}\label{eqn:subset-sum}
        \begin{equation*}
            f=\sum_{i=1}^n x_i-\beta \qedhere
        \end{equation*}
    \end{restatable}    
\end{definition}

When the characteristic of $\mathbb{F}$ is $0$ and $\beta>n$, then the polynomial $f$ is unsatisfiable over the Boolean hypercube (that is $\forall\ \bar{x}\in \set{0,1}^n\ f(\bar{x})\neq 0$). The unsatisfiable system $\set{f=0,X^2 - X=0}$ is known to have a \emph{linear}-$\ips$ refutation computed by a $O(n^3)$ size constant depth formula (ABPs). The following lemma from \cite{FSTW21} gives the upper bound.
 
\begin{lemma}\emph{(\cite[Proposition B1]{FSTW21})}\label{lem:deg-subset-sum}
  Let $\mathbb{F}$ be field of characteristic $0$ and $g\in \mathbb{F}[X]$ be the unique multilinear polynomial such that $g \cdot (\sum_{i=1}^n x_i-\beta) \equiv 1 \mod{X^2 - X}$.
  Then,
  \[
        g = \sum_{i=0}^n \frac{- i!}{\prod_{j=0}^i(\beta-i)} \sum_{S \subseteq [n] : |S|=i}\prod_{i\in S}x_i.
  \]  
\end{lemma}
For any $n,d\in \mathbb{N}$ with $1\leq d\leq n$, we denote the polynomial $\sum_{S\subseteq [n] : |S|=d}\prod_{i\in S} x_i$ by $\esym_{n,d}$,  
which is the \emph{elementary symmetric polynomial}.
The lemma shows that the unique multilinear polynomial $g$  
is a \emph{linear} combination of elementary symmetric polynomials. It is well known from the work of Ben-Or \cite{b83}, that the polynomial $\esym_{n,d}$ can be computed by a depth $3$ formula ($\sum\prod\sum$) of size $O(n^2)$ (see for example \cite{SW01}). 
Since we need to compute every elementary symmetric polynomial of degree $i\in [n]$, the formula complexity of $g$ is $O(n^3)$. 
This clearly shows the $\ips_{\Lin}$ proof complexity of the subset-sum axiom is $O(n^3)$ over characteristic zero fields. 
Moreover, it is well known that $\esym_{n,i}$ can be computed by an ABP of size $O(ni)$. 
Hence, $g$ also has an ABP of size $O(n^3)$. 
Using these, we give the following corollary.

\begin{restatable}{corollary}{fomulaipsub}\label{thm:formula-ips-up}
    Consider the polynomials $f$, $h \in \mathbb{F}[X,Y]$ as described in $\autoref{eq:def-poly2}$ and $\autoref{eq:def-poly1}$ respectively and the unsatisfiable system described in \autoref{thm:sum-roabp-ips-lb}. Then the following holds.

    \begin{itemize}
        \item  If the characteristic of $\mathbb{F}$ is $0$, then there exists a non-multilinear $\ips$ refutation for the unsatisfiable systems $\{f=0,X^2 - X=0,Y^2 - Y=0\}$ and $\{h=0,X^2 - X=0,Y^2 - Y=0\}$, which is computable efficiently by constant depth formulas. 
        \item   Similarly, over field of characteristic $p$, there exists a non-multilinear $\ips$ refutation for the unsatisfiable systems $\{f=0,X^2 - X=0,Y^2 - Y=0\}$ and $\{h=0,X^2 - X=0,Y^2 - Y=0\}$, which is computable by constant depth formulas of size $\poly(n,p)$. 
        \item There is a non-multilinear $\ips$ refutation for the unsatisfiable system given in \autoref{thm:sum-roabp-ips-lb} which has a $\poly(n)$-size ABP, when the characteristic of field is $0$.
    \end{itemize}
\end{restatable}

\begin{proof}
    Since both the polynomials $f$ (\autoref{eq:def-poly2}) and $h$ (\autoref{eq:def-poly1}) are sparse polynomials, we can express them as subset-sum axiom polynomials using a distinct new variable for each distinct monomial. 
    If the sparsity of the polynomial is $s$ (for our case, it is $O(n^2)$), then we will have a subset sum axiom polynomial on $s$ variables.
    In the case of characteristic $0$ fields, we can use the upper bound construction described above to get a $O(s^3)$ size constant depth linear $\ips$ refutation for the new system. 
    Next, we substitute the variables by the monomials to get a non-multilinear constant depth refutation of size $O(n^6)$.  
    We prove this formally for $f$. A similar proof can be shown for $h$.
    
    Recall that $f=f'+1$, where \eqpolytwo*
    Let sparsity of $f'$ be $s=O(\frac{n^2}{\log n})$. 
    Further, let $Z=\set{z_1,\ldots,z_s}$ be a set of variables and define $\tilde{f}(\bar{z})=\sum_{i=1}^s z_i+1$, such that $\tilde{f}(m_1,\ldots,m_s)=f(\bar{x})$ where $m_i\in \supp(f')$. 
    Clearly, $\tilde{f}(\bar{z})$ is unsatisfiable over the Boolean cube. 
    Using \autoref{lem:deg-subset-sum}, there is a multilinear polynomial $\tilde{g}(\bar{z})$ such that $\tilde{g}(\bar{z})\cdot \tilde{f}(\bar{z})\equiv 1 \mod{Z^2-Z}$. 
    That is, there exists polynomials $h_1,\ldots,h_s\in \mathbb{F}[\bar{z}]$ such that the following identity holds,
    \[
        \tilde{g}(\bar{z})\cdot \tilde{f}(\bar{z})+\sum_{i=1}^s h_i(z_i^2-z_i)=1.
    \]
    Substituting back the monomials in place of $z$ variables we get
    \[
        \tilde{g}(\bar{m})\tilde{f}(\bar{m})+\sum_{i=1}^sh_i(\bar{m})(m_i^2-m_i)=1,
    \]
    where $\tilde{m}$ denotes the monomials in $\supp(f)$.
    Therefore, there exist $\set{h'_i}_{i \in [n]}$ such that
    \[
        \tilde{g}(\bar{m})\cdot f(\bar{x})+\sum_{i=1}^n h'(i)(x_i^2-x_i)=1
    \]
    The existences of such $h'_i$ follows from \cite[Claim 3.4]{BLRS25}.
    This equation clearly gives the require upper bound of $O(n^6).$
    Similar arguments work for the unsatisfiable system given in \autoref{thm:sum-roabp-ips-lb}. 
    Here we get a non-multilinear $\ips$ refutation computed by a $\poly(n)$-size ABP.  

    In the case of positive characteristic, we can use the upper bound from the work of Behera, Limaye, Ramanathan, Srinivasan \cite[Theorem 1.8]{BLRS25}. 
\end{proof}

Now we sketch a few details which are noted in the conclusion. 
We start with the following fact, which is given in \cite{FSTW21,HLT24}.
\begin{fact}\label{fact:elem-sym-mult-ips-poly}
Let $n\geq d\in \mathbb{N}$ and $f=\esym_{n,d}-\beta$ for some $\beta\geq \binom{n}{d}\in \mathbb{F}$.
Let $g(\vecx)$ be the unique multilinear polynomial that agrees with $\frac{1}{f}$ over the Boolean hypercube. Then there exists $\beta'\neq 0$ and non-zero field elements $\alpha_d,\alpha_{d+1},\ldots,\alpha_n$ such that $g(\vecx)=\sum_{i=d}^n\alpha_i\esym_{n,i}+\beta'$   
\end{fact}
Since the polynomial $f(\vecx)$ is symmetric and the multilinear polynomial $g(\vecx)$ agrees with $\frac{1}{f(\vecx)}$ over the Boolean hypercube, it must be the case that $g(\vecx)$ is a symmetric multilinear polynomial. 
Hence, it can be expressed as a linear combination of elementary symmetric polynomials $\esym_{n,i}$. 
The fact that the coefficients of $\esym_{n,i}$ for $i < d$ are $0$ follows from \autoref{lem:support-lem-1}. 
Moreover, to the best of our knowledge, the best known formula upper bound for $g(\vecx)$ is $O(dn^2)$. 
This follows by computing each $\esym_{n,i}$ (for every $i\in [d,n]$) by a $O(n^2)$ size formula using Ben-Or's construction \cite{b83}. The circuit complexity of $g(\vecx)$ is $O(n\log^2 n)$ \cite{BCS13} and ABP complexity of $g(\vecx)$ is $O(n^2)$ (using Ben-Or's construction).

\begin{observation}\label{formula-ips-mult-obs}
Consider the polynomial $g(\bar{x})$ defined in \autoref{fact:elem-sym-mult-ips-poly} with $d=\frac{n}{10}$. Then the following statements are true.
\begin{itemize}\itemsep0pt
    \item  Any formula computing a $\multips_{\Lin'}$ refutation of the system $\set{g=0,X^2-X=0}$ must have size at least $\Omega(n^2)$.
    \item  Any ABP computing a $\multips_{\Lin'}$ refutation of the system $\set{g=0,X^2-X=0}$ must have size at least $\Omega(n^2)$.
    \item Any circuit computing a $\multips_{\Lin'}$ refutation of the system $\set{g=0,X^2-X=0}$ must have size least $\Omega(n\log n)$.
\end{itemize}
\end{observation}

\begin{proof}
By \autoref{prop:mult-ips-functional-lb}, if we show a $\Omega(n^2)$ formula (ABP) lower bound for the unique multilinear polynomial $f(\bar{X})$ that agrees with $\frac{1}{g(\bar{X})}$ over the Boolean hypercube, we get our $\multips_{\Lin'}$ refutation lower Bound. By the \autoref{fact:elem-sym-mult-ips-poly}, here $f(\bar{X})=\esym_{n,\frac{n}{10}}$ which has a $\Omega(n^2)$ formula (ABP) lower bound from \cite{CKSV22}. Note that, there is a crucial condition on characteristic of field $\mathbb{F}$, $\Char(\mathbb{F})\nmid n$ in order to hold the lower bound. We do not know yet how to remove this condition. So, it is fair to assume $\Char(\mathbb{F})=0$.
For circuit we refer to the work \cite{BS83}, which shows any circuit computing $\esym_{n,\frac{n}{10}}$ needs size at least $\Omega(n\log n).$ Here also the size of $\mathbb{F}$ need to be either large enough or $\Char(\mathbb{F})=0$ for the lower bound.
\end{proof}

\end{document}